\documentclass[12pt]{article}
\usepackage{epsfig,amssymb,amsmath,psfrag,subfigure,rotate,color}

\allowdisplaybreaks
\newcommand{\insertfig}[2]{\includegraphics[width=#1cm]{#2}}

\def \be  {\begin{equation}}
\def \ee  {\end{equation}}
\def \ba  {\begin{eqnarray}}
\def \ea  {\end{eqnarray}}
\def \baa {\begin{eqnarray*}}
\def \eaa {\end{eqnarray*}}
\def \lab #1 {\label{#1}}

\newcommand\re[1]{(\ref{#1})}
\def\d{\hbox{{d}\kern-.20em\hbox{l}}}

\def \matrix #1 {\left(\begin{array}{cc} #1 \end{array}\right)}

\newcommand \vev [1] {\langle{#1}\rangle}

\newcommand \ket [1] {|{#1}\rangle}
\newcommand \bra [1] {\langle {#1}|}
\newcommand{\bit}[1]{\mbox{\boldmath$#1$}}

\newcommand{\ft}[2]{{\textstyle\frac{#1}{#2}}}
\begin{document}

\begin{titlepage}

\thispagestyle{empty}

\vspace*{3cm}

\centerline{\large \bf  Fermionic pentagons and NMHV hexagon}
\vspace*{1cm}

\centerline{\sc A.V.~Belitsky}

\vspace{10mm}

\centerline{\it Department of Physics, Arizona State University}
\centerline{\it Tempe, AZ 85287-1504, USA}

\vspace{2cm}

\centerline{\bf Abstract}

\vspace{5mm}

We analyze the near-collinear limit of the null polygonal hexagon super Wilson loop in the planar $\mathcal{N} = 4$ superYang-Mills theory. We focus 
on its Grassmann components which are dual to next-to-maximal helicity-violating (NMHV) scattering amplitudes. The kinematics in question is studied 
within a framework of the operator product expansion that encodes propagation of excitations on the background of the color flux tube stretched between 
the sides of Wilson loop contour. While their dispersion relation is known to all orders in 't Hooft coupling from previous studies, we find their form factor 
couplings to the Wilson loop. This is done making use of a particular tessellation of the loop where pentagon transitions play a fundamental role. Being 
interested in NMHV amplitudes, the corresponding building blocks carry a nontrivial charge under the SU(4) R-symmetry group. Restricting the current 
consideration to twist-two accuracy, we analyze two-particle contributions with a fermion as one of the constituents in the pair.  We demonstrate that these 
nonsinglet pentagons obey bootstrap equations that possess consistent solutions for any value of the coupling constant. To confirm the correctness of these 
predictions, we calculate their contribution to the super Wilson loop demonstrating agreement with recent results to four-loop order in 't Hooft coupling.

\end{titlepage}

\setcounter{footnote} 0

\newpage

\pagestyle{plain}
\setcounter{page} 1

{
%\footnotesize 
\tableofcontents}

\newpage

\section{Introduction}

In recent years, the planar maximally supersymmetric gauge theory took on the status of a proverbial ``harmonic oscillator'' of field theories, i.e., a solvable dynamical 
model of gauge interactions in four dimensions. Though the theory is superconformally invariant even on quantum level, it does possess an S-matrix when deformed 
away from four dimensions such that its intrinsic infrared divergences get regularized. So its study is of great interest from the point of view of potentially having valuable 
feedback for realistic theory of particle physics, Quantum Chromodynamics, see, e.g., reviews \cite{Alday:2008yw,Elvang:2013cua,Dixon:2013uaa}.

All on-shell states in $\mathcal{N} = 4$ superYang-Mills theory, i.e., positive and negative gauge bosons $G^\pm$, (anti)gauginos $\Gamma_A, \bar\Gamma_A$ and 
scalars $S_{AB}$, can be assembled into a single superfield \cite{Mandelstam:1982cb,Nair:1988bq} $\Phi (p, \eta) = G^+ (p) + \eta^A \Gamma_A (p) + \ft12 \eta^A \eta^B S_{AB} + \dots$ 
as coefficients accompanying Grassmann variables $\eta^A$ transforming in the fundamental representation of the SU(4) R-symmetry group. As a consequence, the 
scattering superamplitude $\mathcal{A}_n$ of $n$ superparticles $\Phi_i = \Phi (p_i, \eta_i)$ admits a terminating expansion in $\eta$'s. Making use of supersymmetry 
and pulling out the energy-momentum conserving delta function along with the Parke-Taylor denominator \cite{Parke:1986gb}, it reads \cite{Nair:1988bq} 
\begin{align}
\mathcal{A}_n = \frac{i (2\pi)^4 \delta^{4|8} (P)}{\vev{12} \vev{23} \dots \vev{n1}}
\Big(
\mathcal{A}_{n;0} + \mathcal{A}_{n;1} + \dots + \mathcal{A}_{n;n-4} 
\Big)
\, ,
\end{align}
with each term $\mathcal{A}_{n;k}$ being a polynomial of homogeneous degree $4k$ in $\eta$'s. The  $\mathcal{A}_{n;k}$ define N$^{k}$MHV amplitudes.

A deep insight into the structure of the superamplitude was offered by its dual description in terms of the expectation value of the super Wilson loop stretched on a null polygonal
contour\footnote{Though there exists neither a proof of this statement nor a consistent perturbative regularization scheme where the equivalence can be verified
order-by-order in 't Hooft coupling \cite{Belitsky:2011zm}. So to date, it is used as a very inspiring mnemonic.} \cite{Mason:2010yk,CaronHuot:2010ek}. The $n-$site 
super Wilson loop develops a similar truncated series in Grassmann variables
\begin{align}
\mathcal{W}_n =  \mathcal{W}_{n;0} + \mathcal{W}_{n;1} + \dots + \mathcal{W}_{n; n - 4}
\, ,
\end{align}
with each term $\mathcal{W}_{n;k }$ being an SU(4) invariant polynomial possessing a homogeneous Grassmann degree $4k$. The duality between the
super Wilson loop and scattering amplitudes establishes the equality between their expansions as follows
\begin{align}
\mathcal{W}_{n;k} = g^{2k} \mathcal{A}_{n;k}
\, . 
\end{align}

\begin{figure}[t]
\begin{center}
\mbox{
\begin{picture}(0,180)(60,0)
\put(0,-220){\insertfig{18}{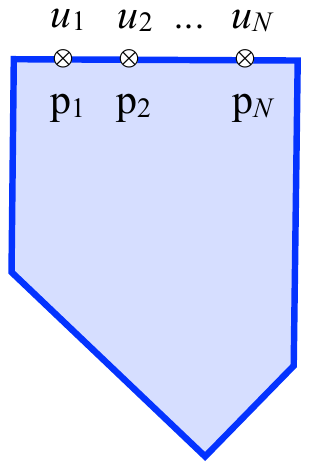}}
\end{picture}
}
\end{center}
\caption{ \label{GenericPentagonFF} Pentagon form factor defining the coupling of flux-tube excitations to the Wilson loop contour.}
\end{figure}

The main advantage of this reformulation is that it provides an opportunity to use dynamics on the two-dimensional world-sheet of the loop with four-dimensional
geometry entering the game only through its boundary \cite{Alday:2007hr}. As a consequence, one can rely on the integrable dynamics of excitations propagating on the 
color flux-tube stretched between a pair of segments of the Wilson loop contour in order to unravel $\mathcal{W}_n$ in a truly nonperturbative manner in 't Hooft coupling. 
A properly constructed finite ratio $\mathcal{W}_n$ of Wilson loop expectation values admits a well-defined expansion in terms of light-ray operators in a given channel 
that is akin to the usual local operator product expansion (OPE) for correlation functions in CFT as was demonstrated in Ref.\ \cite{Alday:2010ku}. The focus of this paper will 
be the six-site superloop, or superhexagon. In particular, we will be after the term in its Grassmann decomposition that corresponds to the NMHV superamplitude. In this 
simplest nontrivial case, the OPE of the Wilson loop receives contribution from a single set of intermediate states such that it reads schematically \cite{Basso:2013vsa}
\begin{align}
\label{GenericOPEffs}
\mathcal{W}_{6;1} = \sum_{N} \int d^N \bit{u} \, F_N (0| \bit{u}) \, {\rm e}^{- \tau E_N (\bit{\scriptstyle u}) + i \sigma p_N  (\bit{\scriptstyle u}) + i \phi m_N} F_N (- \bar{\bit{u}}|0)
\, .
\end{align}
Here in the right-hand side, we suppressed an overall degree-four perfactor in terms of Grassmann variables. The integration in the above equation goes over rapidities of 
excitations $\bit{u} = (u_1, \dots, u_N)$, with a convention introduced for the vector $\bar{\bit{u}} = (u_N, \dots, u_1)$.  Due to the integrable nature of the color flux-tube, 
the $N$-particle energy, momentum and helicity get decomposed into individual single-particle ones
\begin{align}
E_N (\bit{u}) = E_{{\rm p}_1} (u_1) + \dots + E_{{\rm p}_N} (u_N)
\, , \
p_N (\bit{u}) = p_{{\rm p}_1} (u_1) + \dots + p_{{\rm p}_N} (u_N)
\, , \
m_N = m_{{\rm p}_1} + \dots + m_{{\rm p}_N}
\, ,
\end{align}
with subscripts ${\rm p}_i$ designating the type of contributing particles. The fundamental excitations of the flux tube consist of the hole, fermions and gluons (as well 
as bound states of the latter). Their energies and momenta are known nonperturbatively \cite{Basso:2010in}. At vanishing 't Hooft coupling, all 
single-particle energies become degenerate and define the twist of corresponding excitations
\begin{align}
E_{\rm p} (u) |_{g = 0} = 1
\, .
\end{align}
Then it becomes obvious that the expansion \re{GenericOPEffs} receives its leading effect from single-particle states, which scale as ${\rm e}^{- \tau}$, while
the first subleading ${\rm e}^{-2 \tau}$ contribution arises from two particles etc., providing a natural expansion hierarchy. To successfully determine the 
near-collinear expansion of the Wilson loop, one then has to determine the coupling of the flux-tube excitations to the perimeter links. These are encoded in the 
so-called pentagon form factors \cite{Basso:2013vsa}
\begin{align}
F_N (0| \bit{u}) = \bra{ {\rm p}_1(u_1) \dots {\rm p}_N (u_N)} \widehat{\mathcal{P}} \ket{0}
\, ,
\end{align}
shown in Fig.\ \ref{GenericPentagonFF}. These arise from a tessellation of the Wilson loop in terms of the fundamental squares, with pentagons resulting 
from the two adjacent ones  \cite{Basso:2013vsa}. 

%%%%%%%%%%%%%%%%%%%%%%%%%%%%%%%%%%%%%%%%%%%%%%%%%%%%%%%%%%%%%%%%%%%%%
%            Figure
%%%%%%%%%%%%%%%%%%%%%%%%%%%%%%%%%%%%%%%%%%%%%%%%%%%%%%%%%%%%%%%%%%%%%
%%%%%%%%%%%%%%%%%%%%%%%%%%%%%%%%%%%%%%%%%%%%%%%%%%%%%%%%%%%%%%%%%%%%%

In a series of seminal papers  \cite{Basso:2013vsa,Basso:2013aha,Basso:2014koa}, a set of defining equations was proposed and applied to determine pentagon 
form factors with singlet quantum numbers in a given operator channel. In our previous analysis \cite{Belitsky:2014sla}, this formalism was extended to account 
for contributions with non-trivial representations with respect to SU(4), focusing on a specific NMHV channel that transforms in the ${\bf 6}$ of the R-symmetry group. 
Presently, we conclude this discussion by proposing nonperturbative formulas for other two-particle states.

Our subsequent presentation is organized as follows. In the next section, we start exploring pentagon transitions involving a (anti)fermion and a hole. We 
introduce S-matrices with the full SU(4) tensor structure and construct their mirrors in bosonic and fermionic rapidities. Then we introduce a set of axioms, following 
the strategy of Refs.\  \cite{Basso:2013vsa,Basso:2013aha,Basso:2014koa}, for (anti)fermion--hole pentagons and solve them in terms of the ratio of the scattering 
matrix and its mirror up to an overall function that obeys certain crossing and permutation conditions. The bootstrap equations alone do not allow us to constrain 
its form in an unambiguous fashion. We fix the remaining uncertainty in their functional form by confronting chosen ansatze to perturbative data. We provide a similar 
discussion for (anti)fermion--(anti)gluon S-matrices and pentagons in Section \ref{GFsection}. Section \ref{NMHVsection} is dedicated to perturbative tests of our findings 
against available multiloop data for NMHV amplitudes that were made available through recent advances in the hexagon bootstrap approach 
\cite {Dixon:2013eka,Dixon:2014voa,Dixon:2011nj} and allowed to push the current state-of-the-art to four-loop order at the NMHV level \cite{Dixon:2014iba,DixVonHip15}. 
We find a complete agreement. Finally, we conclude. In the appendix, we provide a summary of scattering matrices and their mirrors for all pentagon transitions discussed in 
the main body of the paper.

\section{Hole--fermion pentagon}
\label{HFsection}

We start our discussion with hole--fermion pentagons. In fact, both excitations are charged with respect to the R-symmetry group. Within the asymptotic Bethe Ansatz 
for $\mathcal{N} = 4$ superYang-Mills \cite{Beisert:2005fw}, the SU(4) symmetry gets restored through a particular arrangement of Bethe roots of the momentum carrying 
fermionic roots and isotopic ones that form stacks \cite{Basso:2010in}. The fermion $\Psi^A$ and antifermion $\bar\Psi_A$ flux-tube excitations transform in the fundamental 
${\bf 4}$ and antifundamental ${\bf \bar{4}}$ representation of SU(4), respectively. Analogously, a single hole dual to the central Bethe root gets promoted into a vector of real 
scalars $\Phi_a$ that belong to the representation ${\bf 6}$ of SU(4). We can conveniently recast the latter instead as an antisymmetric rank two tensor of complex scalars 
$\phi^{AB} =  \frac{1}{\sqrt{2}} \Sigma^{AB}_a \Phi_a$ with the help of four-dimensional blocks of six-dimensional Euclidean Dirac matrices. These obey the reality condition 
$\bar\phi_{AB} = (\phi^{AB})^\ast = \ft12 \varepsilon_{ABCD} \phi^{CD}$. The $\Sigma$ and $\bar\Sigma$ blocks obey the Clifford algebra $\bar\Sigma_{a, AC} 
\Sigma_b^{CB} + \bar\Sigma_{b, AC} \Sigma_a^{CB} = 2 \delta_{ab} \delta_A^B$ and are related by complex conjugation $(\Sigma_a^{AB})^\ast = 
- \bar\Sigma_{a,AB} = \bar\Sigma_{a,BA} $. They can be built from 't Hooft symbols $\eta_{iAB}$ and $\bar\eta_{iAB}$ as their elements, i.e.,  $\Sigma_a^{AB} = (i \eta_{iAB}, 
- \bar\eta_{iAB})$ and $\bar\Sigma_{a,AB} = (i \eta_{iAB}, \bar\eta_{iAB})$, see Appendix B of Ref.\ \cite{Belitsky:2003sh}. 

\subsection{Hole--(anti)fermion S-matrix}

The main player in the analysis that follows will be the scattering matrix between (anti)fermions and holes. For a reference state $\ket{{\phi}_{AB} (u) \Psi^C (v)}$,
the S-matrix acts as a permutation operator that interchanges the properly ordered excitations, 
\begin{align}
\ket{
{\phi}_{AB} (u) \Psi^C (v)
}
=
[S_{{\rm h} \Psi} (u, v)]_{AB;D}^{C;EF}
\ket{
 \Psi^D (v) {\phi}_{EF} (u)
}
\, ,
\end{align}
without changing their rapidities. The product of the scalar and fermion in the state can be decomposed into two irreducible components $\bf{6} \otimes \bf{4} = 
\bf{\bar{4}} \oplus \bf{20}$,
\begin{align}
\ket{
{\phi}_{AB} (u) \Psi^C (v)
}
= \left\{
[ \Pi_{\bf{\bar 4}}]^{C;EF}_{AB;D}
+
[ \Pi_{\bf{20}}]^{C;EF}_{AB;D}
\right\}
\ket{
{\phi}_{EF} (u) \Psi^D (v)
}
\, ,
\end{align}
with the help of the projectors
\begin{align}
[ \Pi_{\bf{\bar 4}}]^{C;EF}_{AB;D}
=
\frac{1}{3} \delta_{[B}^C \delta_{A]}^{[E} \delta_D^{F]}
\, , \qquad
[ \Pi_{\bf{20}}]^{C;EF}_{AB;D}
=
\frac{1}{6} \delta_D^{\{C} \delta_H^{G\}}  \varepsilon_{ABGI} \varepsilon^{EFHI}
\, ,
\end{align}
where $[A,B]=AB-BA$ and $(A,B) = AB+BA$ stands for the non-weighted anti- and symmetrization, respectively.
Being projectors, they obey conventional properties
\begin{align}
\label{ProjectorProperties}
[ \Pi_{\bf{r}}]^{C;EF}_{AB;D} [ \Pi_{\bf{r}}]^{D;GH}_{EF;I}  = [ \Pi_{\bf{r}}]^{D;EF}_{AB;I} 
\, , \qquad
[ \Pi_{\bf{r}}]^{C;EF}_{AB;D} [ \Pi_{\bf{r'}}]^{D;GH}_{EF;I}  = 0
\, , \qquad
[ \Pi_{\bf{r}}]^{C;AB}_{AB;C} = \bf{r}
\, .
\end{align} 
Analogously, the hole-fermion S-martix is decomposed in their terms as follows
\begin{align}
[S_{{\rm h} \Psi} (u, v)]_{AB;D}^{C;EF}
=
S_{{\rm h} \Psi} (u, v)
\left\{
\frac{u - v + \ft{3 i}{2}}{u - v - \ft{3 i}{2}}
[ \Pi_{\bf{\bar 4}}]^{C;EF}_{AB;D}
+
[ \Pi_{\bf{20}}]^{C;EF}_{AB;D}
\right\}
\, ,
\end{align}
factoring out a universal SU(4) tensor structure from an overall phase $S_{{\rm h} \Psi} (u, v)$ that is sensitive to the dynamics of the flux tube.
The scattering matrix for the oppositely ordered excitations can be decomposed as
\begin{align}
\label{ScalaPsiTensorSMatrix}
[S_{\Psi {\rm h}} (u, v)]_{AB;D}^{C;EF}
=
S_{\Psi {\rm h}} (u, v)
\left\{
\frac{u - v + \ft{3 i}{2}}{u - v - \ft{3 i}{2}}
[ \Pi_{\bf{\bar 4}}]^{C;EF}_{AB;D}
+
[ \Pi_{\bf{20}}]^{C;EF}_{AB;D}
\right\}
\, ,
\end{align}
such that
\begin{align}
[S_{{\rm h} \Psi} (u, v)]_{AB;D}^{C;EF}
[S_{\Psi {\rm h}} (u, v)]_{EF;H}^{D;IJ}
= \frac{1}{2} \delta^C_H \delta^I_{[A} \delta^J_{B]}
\, .
\end{align}
It is obvious from these definitions that one can introduce scattering matrices in the $\bf{\bar 4}$ and $\bf{20}$ representation of SU(4),
\begin{align}
\label{ChannelSphipsi} 
S^{\bf{\bar 4}}_{{\rm h} \Psi} (u, v) = 
\frac{u - v + \ft{3 i}{2}}{u - v - \ft{3 i}{2}}
S_{{\rm h} \Psi} (u, v)
\, , \qquad
S^{\bf{20}}_{{\rm h} \Psi} (u, v) = 
S_{{\rm h} \Psi} (u, v)
\, .
\end{align}

The explicit nonperturbative form of the overall phase is given in Appendix \ref{ExplicitHFfSmatrices} together with its mirror transform in the hole rapidity, $u
\to u^\gamma = u + i$ along a path that was established in the analysis of Ref.\ \cite{Basso:2011rc} (see Refs.\  \cite{Basso:2013pxa,Belitsky:2014sla} for a detailed 
discussion). The overall phase factor in Eq.\ \re{ScalaPsiTensorSMatrix} obeys the unitarity and crossing conditions,
\begin{align}
\label{SphipsiUnitarity}
S_{{\rm h} \Psi} (u, v)  S_{{\rm h} \Psi} (-u, -v)
= 1
\, , \qquad
S_{{\rm h} \Psi} (u^{2 \gamma}, v) S_{{\rm h} \Psi} (u, v) 
=  
\frac{u - v + \ft{i}{2}}{u - v + \ft{3 i}{2}} 
\, ,
\end{align}
respectively, where in the first relation, we can use instead $S_{\Psi {\rm h}} (v, u) = S_{{\rm h} \Psi} (-u, -v)$.

Let us remind that a (anti)fermion lives on a Riemann surface built from two rapidity $u$-planes glued together along the cut $[-2g, 2g]$ on the real axis \cite{Basso:2010in}. 
On the upper sheet its momentum $p \sim O(1)$ while on the lower sheet $p \sim O(g^2)$ for rapidities $u \sim O(1)$, thus defining large ($\Psi = {\rm F}$) and small 
($\Psi = {\rm f}$) fermion kinematics, respectively. The two are related by an analytic continuation through the above cut \cite{Basso:2010in}. When the small fermion at zero momentum 
comes in a combination with another excitation it acts on it as an operator of supersymmetric transformation \cite{Alday:2007mf}. This will play a pivotal role in the analysis of 
the OPE of the Wilson loop.

\subsection{Anomalous mirror in fermion rapidity}

%%%%%%%%%%%%%%%%%%%%%%%%%%%%%%%%%%%%%%%%%%%%%%%%%%%%%%%%%%%%%%%%%%%%%
%            Figure
%%%%%%%%%%%%%%%%%%%%%%%%%%%%%%%%%%%%%%%%%%%%%%%%%%%%%%%%%%%%%%%%%%%%%
\begin{figure}[t]
\begin{center}
\mbox{
\begin{picture}(0,150)(230,0)
\put(0,-290){\insertfig{20}{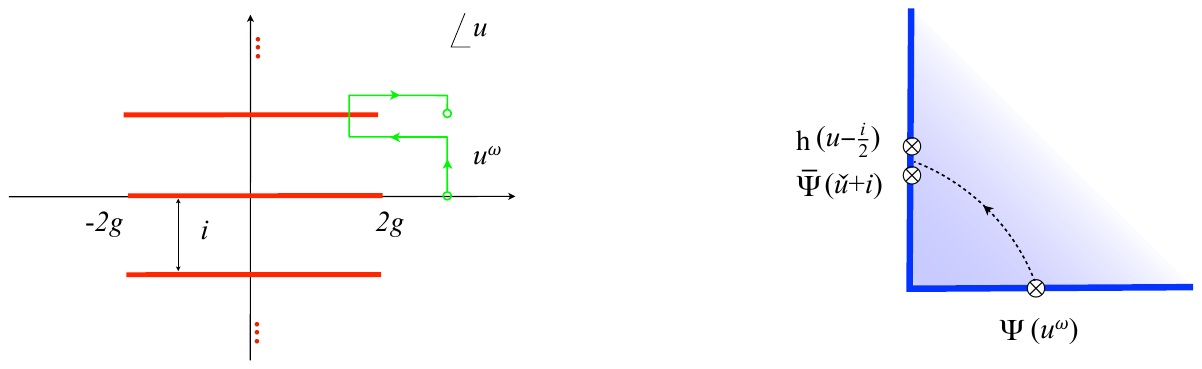}}
\end{picture}
}
\end{center}
\caption{ \label{FermionMirror} The contour for the anomalous mirror transformation of the fermion (left panel) and its graphical representation as the fermion is moved 
into the crossed channel: it becomes a composite state of antifermion on the small sheet (hence its rapidity is dressed with a check $\check{u}$) and a hole (right panel). 
See Ref.\ \cite{Basso:2014koa} for details.}
\end{figure}
%%%%%%%%%%%%%%%%%%%%%%%%%%%%%%%%%%%%%%%%%%%%%%%%%%%%%%%%%%%%%%%%%%%%%

In this section, we will establish the mirror hole--fermion S-matrix when the fermion is moved to the crossed channel. Contrary to the linear nature of the 
transformation for bosonic flux-tube excitations \cite{Basso:2011rc}, the mirror map for the fermion is far more trickier. In fact, as was demonstrated in 
Ref.\ \cite{Basso:2014koa}, a consistent way to mirror the fermion, that sustains a proper change/conservation of its quantum numbers, like 
helicity/R-charge, is achieved by promoting it into a composite antifermion--hole state, see the right panel in Fig.\ \ref{FermionMirror}. This becomes 
obvious from the study of the fermion's dispersion relation when one chooses the path for $\omega$, $u \to u^\omega = u + i$, as shown in Fig.\ 
\ref{FermionMirror} (left panel).

To execute the fermion mirror transformation properly and account for emerging rational factors, we have to work with complete SU(4) tensors.
However, we found it more advantageous to perform the analysis for scalar indices in the vector representation rather than antisymmetric one. That is, we 
contract the above S-matrices with the four-dimensional blocks of six-dimensional Dirac matrices, such that $\ft12 \bar\Sigma_{a,EF} [S_{\rm h {\Psi}}]^{A,EF}_{CD,B} 
\Sigma^{CD}_b = [S_{\rm h {\Psi}}]^{Ab}_{Ba}$. In this notations, the hole--fermion S-matrix reads
\begin{align}
\label{PhiPsiSdirect}
[S_{\Psi{\rm h}} (u, v)]_{Ba}^{Ab} = [R_{\bf 46} (u - v)]_{Ba}^{Ab} S_{\Psi{\rm h}} (u, v)
\, ,
\end{align}
with the R-matrix being \cite{Berg:1977dp,Basso:2014koa}
\begin{align}
[R_{\bf 46} (w)]_{Ba}^{Ab} = \delta_{ab} \delta_B^A + \frac{i}{2w - 3 i}  \Sigma_a^{AC} \bar\Sigma_{b, CB}
\, .
\end{align}
In addition to the fermion--hole S-matrix, we introduce the antifermion--hole one,
\begin{align}
[S_{\bar\Psi {\rm h}} (u, v)]_{Aa}^{Bb} = [R_{\bf \bar{4} 6} (u - v)]_{Aa}^{Bb} S_{\bar\Psi{\rm h}} (u, v)
\, ,
\end{align}
with the SU(4) tensor, in complete analogy with the previous case, being
\begin{align}
\label{R64-bar}
[R_{\bf \bar{4}6} (w)]_{Aa}^{Bb} = \delta_{ab} \delta_A^B + \frac{i}{2w - 3 i} \bar\Sigma_{a, AC}  \Sigma_b^{CB}
\, , 
\end{align}
and an a priori independent phase $S_{\rm {\rm h}\bar\Psi}$. The above SU(4) tensors are related via $[R_{\bf 46} (w)]_{Aa}^{Bb} = [R_{\bf \bar{4}6} (w)]_{Ab}^{Ba}$ 
since ${\bf 6} = {\bf \bar{6}}$ for $SU(4)$. As the hole does not carry spin, it should be indifferent whether it scatters on a fermion or antifermion, which implies that 
the overall phases coincide  \cite{Fioravanti:2013eia}
\begin{align}
\label{HoleFermion=HoleAntifermion}
S_{{\rm h}\bar\Psi} (u, v) = S_{{\rm h}\Psi} (u, v)
\, . 
\end{align}
Next, we recall the index structure of the hole--hole S-matrix
\begin{align}
[S_{{\rm h}{\rm h}} (u, v)]_{ab}^{cd} = [R_{\bf 66} (u - v)]_{ab}^{cd} S_{{\rm h}{\rm h}} (u, v) 
\, ,
\end{align}
that will be involved in the fusion procedure. The $R_{\bf 66}$ tensor coincides with the well-known Zamolodchikov O(6) S-matrix \cite{Zamolodchikov:1977nu}
\begin{align}
[R_{\bf 66} (w)]_{ab}^{cd} 
=
\frac{w}{w - i} \delta_{ac} \delta_{bd} - \frac{i}{w - i} \delta_{ad} \delta_{bc} + \frac{i w}{(w - i)(w - 2i)} \delta_{ab} \delta_{cd}
\, .
\end{align}

%%%%%%%%%%%%%%%%%%%%%%%%%%%%%%%%%%%%%%%%%%%%%%%%%%%%%%%%%%%%%%%%%%%%%
%            Figure
%%%%%%%%%%%%%%%%%%%%%%%%%%%%%%%%%%%%%%%%%%%%%%%%%%%%%%%%%%%%%%%%%%%%%
\begin{figure}[t]
\begin{center}
\mbox{
\begin{picture}(0,160)(250,0)
\put(0,-260){\insertfig{20}{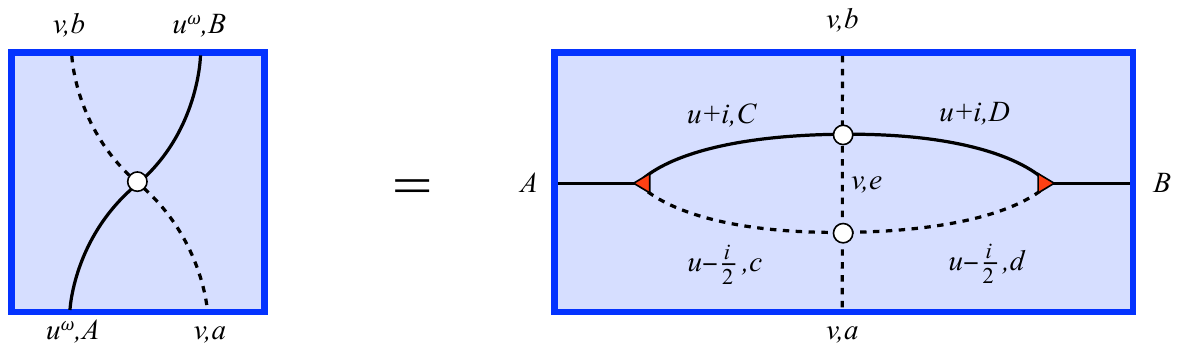}}
\end{picture}
}
\end{center}
\caption{ \label{FusedS} The mirror S-matrix for the fermion and hole as a fusion of the hole-hole and antifermion--hole matrices in the crossed channel.}
\end{figure}
%%%%%%%%%%%%%%%%%%%%%%%%%%%%%%%%%%%%%%%%%%%%%%%%%%%%%%%%%%%%%%%%%%%%%

As in the previously addressed case \cite{Basso:2014koa}, the mirror S-matrix for the fermion in the crossed channel can be found from the anomalous map 
making use of the fusion
\begin{align}
\label{PsiHoleFusion}
[S_{\Psi{\rm h}} (u^\omega, v)]_{Ba}^{Ab}
&
=
S_{\Psi{\rm h}} (u^\omega, v)
[R_{\bf 46} (u - v + i)]_{Ba}^{Ab}
\nonumber\\
&
=
\frac{1}{\sqrt{6}}  \Sigma_c^{AC} 
[S_{\ast\bar\Psi{\rm h}} (\check{u} + i, v)]_{Ce}^{Db}
[S_{\ast{\rm h}{\rm h}} (u - \ft{i}{2}, v)]_{ca}^{de}
\frac{1}{\sqrt{6}} \bar\Sigma_{d,DB} 
\, ,
\end{align}
exhibited in Fig.\ \ref{FusedS}. Here, the right-hand side involves the mirror antifermion--hole and hole--hole S-matrices
\begin{align}
[S_{\ast\bar\Psi {\rm h}} (u, v)]_{Aa}^{Bb}
&
= [R_{\bf \bar{4} 6} (u - v + i)]_{Aa}^{Bb} S_{\ast\bar\Psi{\rm h}} (u, v)
\, , \\
[S_{\ast{\rm h}{\rm h}} (u, v)]_{ab}^{cd} 
&
= [R_{\bf 66} (u - v + i)]_{ab}^{cd} S_{\ast{\rm h}{\rm h}} (u, v) 
\, ,
\end{align}
where the notation $\ast {\rm p}$ implies that the excitation ${\rm p}$ is taken in the crossed channel, e.g., $S_{\ast \rm hh} (u, v) = S_{\rm hh} (u^\gamma, v)$ with 
$u^\gamma$ for holes being the shift $u^\gamma = u + i$ through the cut $[-2g + i/2, 2g + i/2]$ in the complex rapidity plane \cite{Basso:2011rc}. The right-hand 
side of Eq.\ \re{PsiHoleFusion} can be simplified in virtue of the following identity (where $w = u - v + i$)
\begin{align}
\frac{1}{6} \Sigma_c^{AC}  \bar\Sigma_{d,DB} 
[R_{\bf \bar{4}6} (w + i)]_{Ce}^{Bb}
[R_{\bf 66} (w - \ft{i}{2})]_{ca}^{de}
=
\frac{2w + i}{2w - i}
[R_{\bf 46} (w)]_{Ba}^{Ab}
\, ,
\end{align}
and we deduce the definition of the mirror antifermion--hole S-matrix $S_{\ast\bar\Psi{\rm h}}$
\begin{align}
\label{SphiOmegapsi}
S_{\ast\bar\Psi{\rm h}}  (\check{u}+ i, v) S_{\ast{\rm h}{\rm h}} (u - \ft{i}{2}, v) = \frac{u - v + \frac{i}{2}}{u - v + \frac{3 i}{2}} S_{\Psi{\rm h}}  (u^\omega, v)
\, . 
\end{align}
By shifting the $u$-rapidity as $u \to u - i$, we can rewrite the above equation as
\begin{align}
\label{SmirrorPsih}
S_{\ast\bar{\rm f}{\rm h}} (\check{u}, v) = \frac{u - v - \ft{i}{2}}{u - v + \ft{i}{2}} \frac{S_{{\rm Fh}} (u, v)}{S_{{\rm h}{\rm h}} (u - \ft{i}{2}, v)}
\, ,
\end{align}
where we used the fact that $S_{\ast{\rm h}{\rm h}} (u^{- \gamma}, v) = S_{{\rm h}{\rm h}} (u, v)$. The explicit mirror S-matrix can be easily constructed from
this result and its form is deferred to Appendix \ref{HoleFermionMirrorS} where it is displayed for both small and large fermions in the crossed channel.

From explicit diagrammatic representation of the mirror S-matrices, one can establish a chain of relations,
\begin{align}
[S_{\ast{\rm h}\Psi} (u, v)]_{a B}^{b A} = [S_{\ast\bar{\Psi}{\rm h}} (v, u)]_{a B}^{b A} = [S_{\ast\Psi{\rm h}} (v, u)]_{a B}^{b A} 
\, .
\end{align}
Wherefrom we conclude the mirror S-matrix on a fermion or antifermion is the same
\begin{align}
\label{HolePsitoHolePsiBar}
S_{\ast\Psi{\rm h}} (u, v) = S_{\ast\bar{\Psi}{\rm h}} (u, v)
\, ,
\end{align}
while the relation between $S_{\ast{\rm h}\Psi}$ and $S_{\ast\bar{\Psi}{\rm h}}$ is
\begin{align}
\label{SpsiphiRelation} 
S_{\ast\Psi{\rm h}} (u, v) = \frac{u - v - \ft{i}{2}}{u - v + \ft{i}{2}} S_{\ast{\rm h}\bar\Psi} (v, u)
\, .
\end{align}
This identity can be verified using explicit expressions from Appendix \ref{ExplicitHFfSmatrices}.

To find a mirror transformation in both rapidities, we can start  from Eq.\ \re{SphiOmegapsi} and rewrite it in the form
\begin{align}
S_{{\rm h}\Psi} (u, v^\omega) 
=
\frac{u - v - \ft{i}{2}}{u - v - \frac{3i}{2}}
\frac{1}{S_{{\rm h}{\rm h}} (v - \ft{i}{2}, u^{- \gamma}) S_{\Psi{\rm h}} (\check{v} + i, u^{- \gamma})}
\, ,
\end{align}
where we used the mirror transformation for the scalar S-matrix $S_{{\rm h}{\rm h}} (u^\gamma, v^\gamma) = S_{{\rm h}{\rm h}} (u, v)$ \cite{Basso:2013aha}  and 
relations \re{SpsiphiRelation} for $S_{\ast\Psi{\rm h}}$ along with the crossing identity \re{SphipsiUnitarity}. Then performing the mirror transformation of the hole 
rapidity $u$, i.e., $u \to u^\gamma = u + i$, we immediately find
\begin{align}
S_{{\rm h}\Psi} (u^\gamma, v^\omega) 
=
\frac{u - v + \ft{i}{2}}{u - v - \frac{i}{2}}
\frac{1}{S_{{\rm h}{\rm h}} (v - \ft{i}{2}, u) S_{\Psi{\rm h}} (\check{v} + i, u)}
\, .
\end{align}

\subsection{Hole-fermion pentagons and bootstrap}

Having discussed the S-matrices, let us turn to the axioms for the pentagon transitions that are defined by the following matrix elements
\begin{align}
\bra{\Psi (v)} \widehat{\mathcal{P}} \ket{{\rm h} (u)} = P_{{\rm h} | \Psi} (u | v) 
\, , \qquad
\bra{\bar\Psi (v)} \widehat{\mathcal{P}} \ket{{\rm h} (u)} = P_{{\rm h} | \bar\Psi} (u | v) 
\, .
\end{align}
Since the fermion--hole and antifermion--hole S-matrices coincide, see Eq.\ \re{HoleFermion=HoleAntifermion}, the equations which both of the above transitions
obey do coincide. So we will display only one set, with the other one obtained by the substitutions $\Psi \leftrightarrow \bar\Psi$. As advocated at length in the seminal papers 
\cite{Basso:2013vsa,Basso:2013aha}, the bootstrap equations consists from:

\begin{itemize}
\item ``Watson'' equations\footnote{Since these are not the Watson equations in their original incarnation \cite{Watson:1954uc} as the particles belong to the in- and 
out-states, we will use the term in quotes.}:
\begin{align}
\label{hPsiWatsonEq}
P_{{\rm h} | \Psi} (u | v) = S_{{\rm h} \Psi} (u, v) P_{\Psi | {\rm h}} (v | u)
\, , \qquad
P_{\Psi | {\rm h}} (u | v) = S_{\Psi {\rm h}} (u, v) P_{{\rm h} | \Psi} (v | u)
\, ,
\end{align}
\item Mirror equation:
\begin{align}
\label{HoleMirrorBoostrapEq}
P_{{\rm h} | \Psi} (u^{-\gamma}|v) = P_{\bar\Psi | {\rm h}} (v | u)
\, ,
\end{align}
\item Reflection equation:
\begin{align}
P_{{\rm h} | \Psi} (u|v) \sim P_{\bar\Psi | {\rm h}} (-v|-u)
\, ,
\end{align}
\end{itemize}
with $u^\gamma = u + i$ and proportionality factor in the last equation implying presence of a possible relative 
phase. The solution to the axioms can be cast in the form
\begin{align}
\label{PhiPsiSolutions}
P_{{\rm h} | \Psi}^2 (u|v) = w_{{\rm h} \Psi}  (u, v) \frac{S_{{\rm h} \Psi} (u, v)}{S_{{\rm h} \Psi} (u^\gamma, v)}
\, , \qquad
P_{\Psi | {\rm h}}^2 (u|v) = w_{\Psi {\rm h}}  (u, v)  \frac{S_{\Psi {\rm h}} (u, v)}{S_{\Psi {\rm h}} (u, v^{- \gamma})}
\, .
\end{align}
where $w$'s obey the following equation
\begin{align}
\label{whPsiEqs}
\frac{w_{{\rm h} \Psi}  (u^{-\gamma} , v)}{w_{\bar\Psi {\rm h}}  (v, u)} = 1
\, , \qquad
\frac{w_{{\rm h} \Psi}  (u, v)}{w_{\Psi {\rm h}}  (v, u)} = \frac{u - v - \ft{i}{2}}{u - v + \ft{i}{2}} 
\, .
\end{align}
The pentagon amplitudes $P_{{\rm h} | \bar\Psi}$ and $P_{\bar\Psi | {\rm h}}$ are given by the same right-hand sides as in Eqs.\ \re{PhiPsiSolutions}, however, 
with a priori different unknown coefficients $w_{{\rm h} \bar\Psi}$ and $w_{\bar\Psi {\rm h}}$ obeying however identical equations \re{whPsiEqs} with $\Psi 
\leftrightarrow \bar\Psi$. The above two relations can be rewritten for a single function $f_{{\rm h} \Psi} $ upon the substitution
\begin{align}
w_{{\rm h} \Psi}  (u, v) = \frac{f_{{\rm h} \Psi}  (u,v)}{u - v + \ft{i}{2}} 
\, , \qquad 
w_{\Psi {\rm h}}  (u, v) = \frac{f_{{\rm h} \Psi}  (v,u)}{v - u - \ft{i}{2}}
\, ,
\end{align}
and similarly for $\Psi \to \bar\Psi$, where $f$'s obey the crossing relation
\begin{align}
\label{fphiGamma}
f_{{\rm h} \Psi}  (u^{- \gamma}, v) = f_{{\rm h} \bar\Psi}  (u, v)
\, ,
\end{align}
while the reflection identity yields
\begin{align}
\label{fphiReflection}
f_{{\rm h} \Psi} (-u, -v) \sim f_{{\rm h} \bar\Psi}  (u, v)
\, ,
\end{align}
up to a phase.
These equations can be solved with a function that is independent of the rapidity $u$, i.e., $f_{{\rm h} \Psi/\bar\Psi}  (u, v) = f_{{\rm h} \Psi/\bar\Psi}  (v)$.
The dependence on $v$ has to be taken in the form of a power of the Zhukowski variable
\begin{align}
x [u] = \ft12 \big( u + \sqrt{u^2 - (2g)^2} \big)
\, ,
\end{align}
such that
\begin{align}
\label{fhPsiexplicit}
f_{{\rm h}\Psi} (v) = f_{{\rm h}\bar\Psi} (v) = x^\alpha[v]
\, ,
\end{align}
and provides agreement with multiloop data as argued below for the choice $\alpha = 0$.

\subsection{Form factors}

With all pentagons fixed, we now transform them into form factors. As all scattering matrices in this sector, the latter can be decomposed
in terms of irreducible components such that they obey individual Watson equations,
\begin{align}
F^{\bf{r}}_{{\rm h}\Psi} (0|u, v) = S^{\bf{r}}_{\Psi{\rm h}} (v, u) F^{\bf{r}}_{\Psi{\rm h}} (0|v, u)
\, , \qquad 
F^{\bf{r}}_{\Psi{\rm h}} (0|u, v) = S^{\bf{r}}_{{\rm h}\Psi} (v, u) F^{\bf{r}}_{{\rm h}\Psi} (0|v, u)
\end{align}
with $\bf{r} = \bf{\bar{4}}, \bf{20}$. Using the explicit form of the scattering matrices \re{ChannelSphipsi}, we immediately conclude
\begin{align}
F^{\bf{20}}_{{\rm h}\Psi} (0|u, v) = (u - v + \ft{3 i}{2}) F^{\bf{\bar{4}}}_{{\rm h}\Psi} (0|u, v)
\, , \qquad
F^{\bf{20}}_{\Psi{\rm h}} (0|u, v) = (v - u - \ft{3 i}{2}) F^{\bf{\bar{4}}}_{\Psi{\rm h}} (0|u, v)
\end{align}
up to a phase. Reflection and cyclic symmetry of the pentagon yield
\begin{align}
F^{\bf{\bar{4}}}_{{\rm h}\Psi} (0|u, v) \sim F^{\bf{\bar{4}}}_{\Psi{\rm h}} (0|-v, -u)
\, , \qquad 
F^{\bf{20}}_{{\rm h}\Psi} (0|u, v) \sim F^{\bf{20}}_{\Psi{\rm h}} (0|-v, -u)
\, ,
\end{align}
again, up to a phase.
The form factors in the ${\bf \bar 4}$ are then obtained by performing multiple mirror transformations on the pentagons, i.e., 
\begin{align}
F^{\bf{\bar 4}}_{{\rm h} \Psi} (0|u, v) = P_{{\rm h} | \Psi} (u^{2\gamma}|v)
\, , \qquad
F^{\bf{\bar 4}}_{\Psi {\rm h}} (0|u, v) = P_{{\rm h} | \Psi} (v^{-3 \gamma} | u)
\, .
\end{align}
The consistency of these equations with the ``Watson" equations \re{hPsiWatsonEq} can be easily verified as consequence of the relation
\begin{align}
S^{\bf{\bar 4}}_{{\rm h}\Psi} (u, v) = \frac{F^{\bf{\bar 4}}_{\Psi {\rm h}} (0|v, u)}{F^{\bf{\bar 4}}_{{\rm h} \Psi} (0|u, v)}
=
\frac{P_{{\rm h} | \Psi} (u^{- 3 \gamma}| v)}{P_{{\rm h} | \Psi} (u^{2 \gamma}| v)}
=
\frac{
S_{\Psi{\rm h}} (v, u^{- 2 \gamma}) 
S_{\Psi{\rm h}} (v, u^{- \gamma}) 
S_{\Psi{\rm h}} (v, u) 
}{
S_{{\rm h}\Psi} (u^{2 \gamma}, v) 
S_{{\rm h}\Psi} (u^{\gamma}, v) 
}
\, ,
\end{align}
which follows from the unitarity and crossing properties of the S-matrix \re{SphipsiUnitarity}.

Using the explicit solutions \re{PhiPsiSolutions} to bootstrap equations, we can find the two-particle form factors in terms of pentagons as
\begin{align}
F^{\bf{\bar 4}}_{{\rm h} \Psi} (0|u, v) 
= \frac{1}{(u - v + \frac{3i}{2}) P_{{\rm h} \Psi} (u|v)}
\, , \qquad
F^{\bf{20}}_{{\rm h} | \Psi} (0|u, v) 
= \frac{1}{P_{{\rm h} | \Psi} (u|v)}
\, ,
\end{align}
and anologously
\begin{align}
\label{HoleAntiFff4}
F^{\bf{4}}_{ {\rm h} \bar \Psi} (0|u, v) 
= 
\frac{1}{(u - v + \frac{3i}{2}) P_{ {\rm h} | \bar\Psi} (u|v)}
\, , \qquad
F^{\bf{20}}_{ {\rm h} \bar \Psi} (0|u, v) 
= 
\frac{1}{P_{ {\rm h} | \bar\Psi} (u|v)}
\, .
\end{align}

\section{Gluon--fermion pentagons}
\label{GFsection}

Next, we turn to the derivation of the pentagon form factors involving (anti)fermion and gauge fields.  The gauge fields come in helicity plus and minus 
eigenstates and will be dubbed correspondingly as gluon and antigluon in what follows.

\subsection{(Anti)gluon--(anti)fermion S-matrix}

Since the gluon is not charged under the R-symmetry group, the SU(4) index structure of the gluon--fermion and gluon--antifermion S-matrices is trivial
\begin{align}
[S_{\rm g {\Psi}} (u, v)]_B^A = \delta_B^A S_{\rm g {\Psi}} (u, v)
\, , \qquad
[S_{\rm g {\bar\Psi}} (u, v)]_A^B = \delta_A^B S_{\rm g {\bar\Psi}} (u, v)
\, .
\end{align}
The scalar phase factors accompanying the above SU(4) Kronecker symbols are not independent from each other and in fact are equal up to an overall rational 
factor \cite{Fioravanti:2013eia}
\begin{align}
\label{SgbarpsiToSgpsi}
S_{\rm g {\bar\Psi}} (u, v) = \frac{u - v + \ft{i}{2}}{u - v - \ft{i}{2}} S_{\rm g {\Psi}} (u, v)
\, .
\end{align}
By helicity conservation, we can also establish the relations
\begin{align}
\label{SgbarpsiToSgpsi2}
S_{\rm g {\bar\Psi}} (u, v) = S_{\rm \bar{g} {\Psi}} (u, v)
\, , \qquad
S_{\rm g {\Psi}} (u, v) = S_{\rm \bar{g} {\bar\Psi}} (u, v)
\, . 
\end{align}
Thus, the only independent dynamical phase $S_{\rm g {\Psi}} (u, v)$, whose  expression is given in Appendix \ref{GaugeFermionSmatrix}, can be found to obey the 
unitarity and crossing conditions
\begin{align}
\label{gPsiMirror}
S_{\rm g {\Psi}} (u, v) S_{\rm g {\Psi}} (- u, - v) = 1
\, , \qquad
S_{\rm g {\Psi}} (u^{2\gamma}, v) S_{\rm g {\Psi}} (u, v) = \frac{u - v -\ft{i}{2}}{u - v + \ft{i}{2}} 
\, ,
\end{align}
where the path to the mirror sheet in the gluon rapidity $u \to u^\gamma = u$ was elaborated in Ref.\ \cite {Basso:2011rc} (see also \cite{Belitsky:2014sla}, for examples 
worked out in great detail). Its explicit form is quoted in Appendix \ref{GaugeFermionSmatrix}. Notice that $S_{{\Psi} \rm g } (v, u) = S_{\rm g {\Psi}} (- u, - v) $ and 
one can also fix $S_{\bar\Psi g} (u, v)$ to be
\begin{align}
\label{SgbarpsiToSgpsi3}
S_{{\bar\Psi} \rm g} (u, v) = \frac{u - v + \ft{i}{2}}{u - v - \ft{i}{2}} S_{{\Psi}\rm g} (u, v)
\, .
\end{align}

\subsection{Anomalous mirror}

The previously discussed anomalous map for the fermions implies that one can construct the mirror fermion--gluon S-matrix $S_{\ast{\Psi}\rm g}$ again by a fusion procedure.
Performing this transformation in the fermion rapidity $u \to u^\omega = u + i$, on the large-fermion--gluon S-matrix from Appendix \ref{GaugeFermionSmatrix}, we immediately 
conclude that
\begin{align}
\label{SFomegaG}
S_{\rm Fg} (u^\omega, v)
=
S_{\ast \rm hg} (u - \ft{i}{2}, v) S_{\ast\rm \bar{f}g} (u + i, v)
\, .
\end{align}
Shifting the rapidity back $u \to u - i$ in the above equation, we obtain
\begin{align}
\label{fbargmirrorSmatrix}
S_{\ast\rm \bar{f}g} (u, v)
=
\frac{S_{\rm Fg} (u, v)}{S_{\rm hg} (u - \ft{i}{2}, v) }
\, .
\end{align}
This equation produces the result quoted in Appendix \ref{GaugeFermionMirror}. By moving the small fermion rapidity to the large sheet, we get the $S_{\ast\rm \bar{F}g} (u, v)$, 
that together with $S_{\ast\rm \bar{f}g} (u, v)$ will be cumulatively denoted by $S_{\ast\rm \bar{\Psi}g} (u, v)$. In order to find the mirror fermion--gluon S-matrix, we reply on the 
relation \re{SgbarpsiToSgpsi} and deduce
\begin{align}
S_{\ast{\Psi} \rm g} (u, v) = \frac{u - v + \ft{i}{2}}{u - v - \ft{i}{2}} S_{\ast{\bar\Psi} \rm g} (u, v) 
\, ,
\end{align}
for either the small or large fermion $\Psi = {\rm f, F}$. 

Let us finally point out that Eq.\ \re{SFomegaG} can be rewritten in the following generic form\footnote{We remind that the check on top of the corresponding fermionic rapidity 
$\check{v}$ implies that it resides on the small sheet.}
\begin{align}
\label{PsiOmegaGluon}
S_{\Psi \rm g} (v^\omega, u) = \frac{1}{S_{\rm gh} (u^{- \gamma}, v - \ft{i}{2}) S_{\rm g {\Psi}} (u^{- \gamma}, \check{v} + i)}
\, , 
\end{align}
with its right-hand side involving only bosons in the mirror channel. This was achieved by means of the following relations between the mirror scattering matrices
\begin{align}
S_{\rm hg} (v^\gamma, u) = S_{\rm hg} (v, u^{- \gamma}) 
\, , \qquad
S_{\ast\bar\Psi \rm g} (v, u) = S_{\ast {\rm g} \bar\Psi} (u, v)
\, ,
\end{align}
and Eq.\ \re{gPsiMirror} applied to the last identity. Finally, mirror transforming the gluon rapidity $u \to u^\gamma$ in Eq.\ \re{PsiOmegaGluon}, 
we immediately conclude that
\begin{align}
S_{\Psi \rm g} (v^\omega, u^\gamma) = \frac{1}{S_{\rm gh} (u, v - \ft{i}{2}) S_{\rm g {\Psi}} (u, \check{v} + i)}
\, ,
\end{align}
completing the list of mirror transformations in both rapidities.

\subsection{Gluon--fermion pentagons}

Having found the explicit gluon--fermion S-matrices and their mirrors in both gluon and fermion flux-tube rapidities, we can now proceed with the construction of  
corresponding pentagons. We can introduce the following fermion--(anti)gluon pentagons, as matrix elements of the pentagon operator between corresponding 
states of the flux tube
\begin{align}
P_{\rm g | {\Psi}} (u|v) 
&= \bra{{\Psi} (v)} \widehat{\mathcal{P}} \ket{{\rm g}(u)}
\, , \qquad
P_{\rm \bar{g} | {\Psi}} (u|v) = \bra{{\Psi} (v)} \widehat{\mathcal{P}} \ket{{\rm \bar{g}}(u)}
\, , \\
P_{{\Psi} | \rm g} (u|v) 
&= \bra{{\rm g} (v)} \widehat{\mathcal{P}} \ket{{\Psi} (u)}
\, , \qquad
P_{{\Psi} | \rm \bar{g}} (u|v) = \bra{{\rm \bar{g}} (v)} \widehat{\mathcal{P}} \ket{{\Psi} (u)}
\, .
\end{align}
The rest, i.e., pentagons involving $\bar\Psi$, can be obtained from these via the equations
\begin{align}
P_{\rm \bar{g} | {\Psi}} (u|v) = P_{\rm g | {\bar\Psi}} (u|v) 
\, , \qquad
P_{\rm g | {\Psi}} (u|v) 
=
P_{\rm \bar{g} | {\bar\Psi}} (u|v) 
\, ,
\end{align}
etc. The pentagon transitions obey a set of axioms that fix them almost
uniquely. For the case at hand, the defining equations take the form
\begin{itemize}
\item ``Watson'' equations:
\begin{align}
\label{gPsiWatson}
P_{\rm g | {\Psi}} (u|v) = S_{\rm g {\Psi}} (u, v) P_{\Psi | \rm g} (v|u)
\, , \qquad
P_{\rm \bar{g} | {\Psi}} (u|v) = S_{\rm \bar{g} {\Psi}} (u, v) P_{\Psi | \rm \bar{g}} (v|u) 
\end{align}
\item Mirror equations:
\begin{align}
\label{BosonMirrorPentagon}
P_{\rm g | {\Psi}} (u^{- \gamma}|v) = P_{{\Psi} | \rm \bar{g}} (v|u)
\, , \qquad
P_{{\Psi} | \rm g} (u|v^{\gamma}) = P_{\rm \bar{g} | {\Psi}} (v|u)
\, .
\end{align}
\item Reflection equations:
\begin{align}
P_{{\Psi} | \rm g} (u|v) \sim P_{\rm g | {\Psi}} (-v|-u)
\, , \qquad
P_{{\Psi} | \rm \bar{g}} (u|v) \sim P_{\rm \bar{g} | {\Psi}} (-v|-u)
\, .
\end{align}
\end{itemize}
All pentagon transitions admit the following universal structure 
\begin{align}
\label{gpsi1Ansatz}
P^2_{\rm g | {\Psi}} (u|v) 
&= w_{\rm g {\Psi}} (u, v) \frac{S_{\rm g {\Psi}} (u, v)}{S_{\rm g {\Psi}} (u^\gamma, v)}
\, , \qquad \ \
P^2_{\rm \bar{g} | {\Psi}} (u|v) = w_{\rm \bar{g} {\Psi}} (u, v) \frac{S_{\rm \bar{g} {\Psi}} (u, v)}{S_{\rm \bar{g} {\Psi}} (u^\gamma, v)}
\, , \\
\label{gpsi2Ansatz}
P^2_{{\Psi} | \rm g} (u|v) 
&= w_{{\Psi}\rm g} (u, v) \frac{S_{{\Psi}\rm g} (u, v)}{S_{{\Psi}\rm g} (u, v^{- \gamma})}
\, , \qquad
P^2_{{\Psi} | \rm \bar{g}} (u|v) = w_{{\Psi}\rm \bar{g}} (u, v) \frac{S_{{\Psi}\rm \bar{g}} (u, v)}{S_{{\Psi}\rm \bar{g}}  (u, v^{- \gamma})}
\, ,
\end{align}
in terms of the S-matrices and their mirrors in bosonic rapidities (see Appendix \ref{GaugeFermionMirror}) up to yet to be determined
functions $w$'s. Due to the relation between the scattering matrices \re{SgbarpsiToSgpsi2}, we can relate pentagons involving negative
and positive-helicity gluons,
\begin{align}
\label{GFandGbarFrelations}
w_{\rm g {\Psi}} (u, v) P^2_{\rm \bar{g} | {\Psi}} (u|v) = w_{\rm \bar{g} {\Psi}} (u, v) P^2_{\rm g | {\Psi}} (u|v) 
\, , \qquad
w_{{\Psi}\rm g} (u, v) P^2_{{\Psi} | \rm \bar{g}} (u|v) = w_{{\Psi}\rm \bar{g}} (u, v) P^2_{{\Psi} | \rm g} (u|v) 
\, .
\end{align}
Substituting the ansatze \re{gpsi1Ansatz} and \re{gpsi2Ansatz} into the axiom equations, we deduce the following relations between the coefficient functions $w$'s,
\begin{align}
\frac{w_{\rm g {\Psi}} (u, v)}{w_{{\Psi}\rm g} (v, u)} = \frac{u - v - \frac{i}{2}}{u - v + \frac{i}{2}}
\, , \qquad
\frac{w_{\rm \bar{g} {\Psi}} (u, v)}{w_{{\Psi}\rm \bar{g}} (v, u)} = \frac{u - v + \frac{i}{2}}{u - v - \frac{i}{2}}
\, ,
\end{align}
and
\begin{align}
\label{GFmirror}
w_{\rm g {\Psi}} (u^{- \gamma}, v) = w_{{\Psi} \rm \bar{g}} (v, u)
\, , \qquad
w_{\rm \bar{g} {\Psi}} (u^{- \gamma}, v) = w_{{\Psi} \rm g} (v, u)
\, ,
\end{align}
together with
\begin{align}
w_{{\Psi}\rm g} (u, v) \sim w_{\rm g {\Psi}} (-v, -u)
\, , \qquad
w_{{\Psi}\rm \bar{g}} (u, v) \sim w_{\rm \bar{g} {\Psi}} (-v, -u)
\, ,
\end{align}
with the equality holding up to a phase.

The solution to these equations is ambiguous. Below we present the one that correctly reproduces the low-loop data (up to four loops), when expanded in 
perturbative series. Notice that once the form is fixed at lowest orders, the bootstrap to nonperturbative dependence in $g^2$ is unique. First, we factor out a 
rational prefactor,
\begin{align}
\label{wgPsi}
w_{\rm g {\Psi}} (u, v) = f_{\rm g{\Psi}} (u, v) (u - v + \ft{i}{2} )
\, , \qquad
w_{{\Psi}\rm g} (u, v) = - f_{\rm g{\Psi}}  (v, u) (u - v + \ft{i}{2} )
\, ,
\end{align}
and similarly
\begin{align}
\label{wgbarPsi}
w_{\rm \bar{g} {\Psi}} (u, v) = \frac{\bar{f}_{\rm {\bar g}{\Psi}} (u, v)}{u - v + \ft{i}{2}}
\, , \qquad
w_{{\Psi}\rm {\bar g}} (u, v) = - \frac{\bar{f}_{\rm {\bar g}{\Psi}} (v, u)}{u - v + \ft{i}{2}}
\, .
\end{align}
Substituting these in the mirror identities \re{GFmirror}, we find that the residual functions $f$ and $\bar{f}$ obey the relation
\begin{align}
\label{ffbarMirror}
\frac{\bar{f}_{\rm \bar{g}{\Psi}}  (u^{-\gamma}, v)}{f_{\rm g{\Psi}}  (u, v)} = (u - v - \ft{i}{2})(u - v + \ft{i}{2})
\, .
\end{align}
This can be solved, for the large fermion ${\Psi} = {\rm F}$, with
\begin{align}
\label{fF}
\bar{f}_{\rm {\bar g}F} (u, v) 
=
[ f_{\rm gF}  (u, v) ]^{-1}
=
\frac{
\left( x^+[u] - x[v] \right) \left(x^-[u] - x[v]  \right)}{x[v]}  \, ,
\end{align}
with adopted conventional notations $x^\pm [u] \equiv x [u^\pm]$ where $u^\pm = u \pm \ft{i}{2}$. 
It is important to realize that \re{ffbarMirror} determines the right-hand side of Eq.\ \re{fF} up to a product of functions depending on corresponding rapidities,
$G_{\rm g} (u) G_{\Psi} (v)$ and $\bar{G}_{\rm g} (u) \bar{G}_{\rm \Psi} (v)$ for $f$ and $\bar{f}$, respectively. While the mirror transformation for the gluon 
flux-tube excitations suggests that $\bar{G}_{\rm g} (u^{- \gamma}) = G_{\rm g} (u)$, the equation relating $G_{\Psi} (v)$ and $\bar{G}_{\Psi} (v)$ should 
be fixed from the mirror transformation involving the fermion. In both cases, the simplest solution $G_{\rm g} = \dots = 1$ will provide agreement with
data.

For the small fermion, we just have to pass in the above formulas to the small fermion sheet by means of an analytic continuation \cite{Basso:2010in}. While
the corresponding S-matrices were introduced earlier in Eqs.\ \re{Sgf} and \re{Sgfmirror}, the passage to the small fermion implies the substitution 
$x[v] \to g^2/x[v]$ in the $f$ and $\bar{f}$ functions \re{fF}, such that they read
\begin{align}
\label{ff}
\bar{f}_{\rm {\bar g}f} (u, v) 
=
[f_{\rm gf}  (u, v) ]^{-1}
=
\frac{x[v] x^+[u] x^-[u]}{g^2}
\left( 1 - \frac{g^2}{ x[v] x^+[u]} \right) \left(1 - \frac{g^2}{ x[v] x^-[u]} \right)
\, .
\end{align}

\subsection{Form factors}

Having found the pentagon transitions, we can derive the pentagon form factors, where all excitations belong to the same side (see
Fig.\ \ref{GenericPentagonFF}) by moving excitations by means of a double mirror \cite{Basso:2013vsa}. Relying on the explicit form of the deduced solutions, we
get the form factor coupling of fermion--(anti)gluon excitations to the Wilson loop contour 
\begin{align}
F_{\rm g {\Psi}} (0|u, v) \equiv P_{\rm {\bar g} | {\Psi}} (u^{2 \gamma}|v) = \frac{1}{P_{\rm g | {\Psi}} (u|v)}
\, , \qquad
F_{\rm \bar{g} {\Psi}} (0|u, v) \equiv P_{\rm g | {\Psi}} (u^{2 \gamma}|v) = \frac{1}{P_{\rm {\bar g} | {\Psi}} (u|v)}
\, .
\end{align}
Here we relied on the fact that
\begin{align}
w_{\rm {\bar g} {\Psi}} (u^{2 \gamma}, v) w_{\rm g {\Psi}} (u, v) = 1
\, .
\end{align}

\section{OPE for NMHV hexagon}
\label{NMHVsection}

The preceding two sections summarized our analysis of two-particle form factors which define the coupling of the flux-tube excitations to the Wilson loop contour within 
the formalism of the operator product expansion. To  compare the super Wilson loop observable derived from the OPE to the six-particle NMHV scattering amplitude we 
have to construct the following combination
\begin{align}
\mathcal{W}_{6;1} = {\mathcal P}_6 W_6
\, ,
\end{align}
where ${\mathcal P}_6$ is the ratio of six-particle superamplitude to its bosonic cousin while $W_6$ is a properly subtracted bosonic hexagon 
\cite{Alday:2010ku}. The former factor admits the representation \cite{Drummond:2008vq}
\begin{align}
{\mathcal P}_6 = \frac{{\mathcal A}_{6;1}}{\mathcal{A}_{6;0}}
&
=
[(2) + (5)] V (u, v, w; g) + [(3) + (6)] V (v, w, u; g) + [(1) + (4)] V (w, u, v; g) 
\nonumber\\
&
-
[(2) - (5)] \widetilde{V} (u, v, w; g) + [(3) - (6)] \widetilde{V} (v, w, u; g) + [(1) - (4)] \widetilde{V} (w, u, v; g) 
\, ,
\end{align}
in terms of superconformal invariants \cite{Drummond:2008vq,Mason:2009qx} defined by a five-bracket, e.g., 
\begin{align}
(1) \equiv [23456] = \frac{\delta^{0|4} \left(\chi_2 (3456) + \chi_3 (4562) + \chi_4 (5623) + \chi_5 (6234) + \chi_6 (2345) \right)}{(2345)(3456)(4562)(5623)(6234)}
\, ,
\end{align}
which are built out of momentum twistors $Z_i^A$, $(ijkl) = \varepsilon_{ABCD} Z_i^A Z_j^B Z_k^C Z_l^D$. These are accompanied by the functions  $V$ and $\widetilde{V}$ 
of three conformal cross-ratios $u,v,w$ and 't Hooft coupling $g$. They admit perturbative expansion\footnote{Notice that due to different normalization of the 
't Hooft coupling the hexagon function differ by a numerical factor from $V^{(\ell)}_{\rm DvH}$ and $\widetilde{V}^{(\ell)}_{\rm DvH}$ introduced in Ref.\ \cite{Dixon:2014iba}. 
Namely, $V^{(\ell)}= 2^\ell V^{(\ell)}_{\rm DvH}$ and correspondingly $\widetilde{V}^{(\ell)} = 2^\ell \widetilde{V}^{(\ell)}_{\rm DvH} $. Also it is important to realize that the 
conformal invariants we use here are related to those in \cite{Dixon:2014iba} by a cyclic permutation $u = v_{\rm DvH}$, $v = w_{\rm DvH}$ and $w = u_{\rm DvH}$.}  in $g$ 
\begin{align}
V (u, v, w; g) = 1 + \sum_{\ell \geq 1} g^{2 \ell} V^{(\ell)} (u, v, w)
\, , \qquad
\widetilde{V} (u, v, w; g) = \sum_{\ell \geq 2} g^{2 \ell} \widetilde{V}^{(\ell)} (u, v, w)
\, .
\end{align}
These functions were recently computed within the so-called hexagon bootstrap program to four-loop order in \cite{Dixon:2014iba,DixVonHip15}, generalizing an earlier two-loop 
consideration of Ref.\ \cite{Dixon:2011nj}. The bosonic hexagon observable, on the other hand, is determined by the product 
\begin{align}
W_6 = W_6^{\rm U(1)} \exp \left( R_6 \right)
\end{align}
of the ratio of the Wilson loops computed in U(1) theory \cite{Alday:2010ku} with the coupling constant determined by the cusp anomalous dimension
\begin{align}
W_6^{\rm U(1)}  = \exp \left[ \ft{1}{4} \Gamma_{\rm cusp} (g) X_6 (u, v, w) \right]
\, ,
\end{align}
where the dependence on cross-ratios is encoded into the function \cite{Gaiotto:2011dt}
\begin{align}
X_6 (u, v, w) = - {\rm Li}_2 (1-u) - {\rm Li}_2 (1-v) - {\rm Li}_2 (1-w) - \ln u \ln w + \ln(1-v) \ln \frac{(1-v)u}{v w}+ 2 \zeta_2
\, ,
\end{align}
and the remainder function $R_6$ of the bosonic hexagon that was determined up to three loop order\footnote{Again, due to difference of 't Hooft couplings, the 
perturbative expansion for the remainder function $R_6$ reads in terms of two-loop $R_6^{(2)}$, originally calculated in Ref.\ \cite{DelDuca:2010zg} and simplified 
making use of the symbol technology in Ref.\ \cite{Goncharov:2010jf}, and three-loop $R_6^{(3)}$ results of Ref.\ \cite{Dixon:2013eka} $R_6 = 4 g^4 R_6^{(2)} + 8 
g^6 R_6^{(3)} + \dots$.} in Ref.\ \cite{Dixon:2013eka}.

To test all fermonic pentagons discussed above, it suffices to extract the $\chi_1^3 \chi_4$ component of the NMHV amplitude. It receives the contribution from 
states transforming under ${\bf 4}$ of SU(4) such that $\mathcal{W}_{6;1}$ admits the following structure in the operator product expansion
\begin{align}
\mathcal{W}_{6;1} 
= 
\chi_1^3 \chi_4 
\left( {\rm e}^{- \tau} {\rm e}^{i \phi/2} \, W_{\Psi}^{(1)} + {\rm e}^{-2 \tau} {\rm e}^{3 i \phi/2} \, W_{\rm g{\Psi}}^{(2)}
+ {\rm e}^{-2 \tau} {\rm e}^{-i \phi/2} \, W_{\rm h {\bar\Psi}/\bar{g} {\Psi}}^{(2)} 
+ O({\rm e}^{-3 \tau}) \right)
+ \dots
\, ,
\end{align}
for the bosonic twistors that parametrize the hexagon taken in the form 
\begin{align*}
Z_1 
&= ({\rm e}^{\sigma - i \varphi/2}, 0, {\rm e}^{\tau + i \varphi/2}, {\rm e}^{-\tau+ i \varphi/2})
\, , \quad
Z_2 
= (1, 0, 0, 0)
\, , \quad
Z_3 
= (-1, 0, 0, 1)
\, , \\
Z_4 
&= (0, 1, -1, 1)
\, , \qquad\qquad\qquad\qquad
Z_5 
= (0, 1, 0, 0)
\, , \quad
Z_6 
= (0, {\rm e}^{- \sigma - i \varphi/2}, {\rm e}^{\tau + i \varphi/2}, 0)
\, .
\end{align*}
Here, the individual twist-$n$ contributions $W^{(n)}$ are functions of the variables $\sigma$ and $\tau$ and the coupling constant $g$. Their particle content 
is displayed as subscripts realized on the basis of their total helicity and R-charge. In perturbation theory, the $\tau$-dependence of $W^{(n)}$ is 
polynomial of order $\ell$ for $O (g^{2\ell})$, while the $\sigma$-dependence arises as nontrivial functions which can be expressed in terms of harmonic 
polylogarithms \cite{Remiddi:1999ew,Maitre:2005uu} and values of the zeta function. The contribution of the Grassmann component $\chi_1^3 \chi_4$ to 
the superconformal invariants, reads to order ${\rm e}^{-2 \tau}$,
\begin{align}
(2)+(5) 
&
= \chi_1^3 \chi_4 \left[  
{\rm e}^{- \tau}  \frac{1 - {\rm e}^{2 \sigma}}{1 + {\rm e}^{2 \sigma}} {\rm e}^{i \phi/2} 
-
{\rm e}^{-2 \tau}
\frac{{\rm e}^\sigma - 2 {\rm e}^{3 \sigma} - {\rm e}^{5 \sigma}}{(1 + {\rm e}^{2 \sigma})^2}
 {\rm e}^{3i \phi/2 } 
-
{\rm e}^{- 2 \tau}  \frac{2 {\rm e}^\sigma + 4 {\rm e}^{3 \sigma}}{(1 + {\rm e}^{2 \sigma})^2} {\rm e}^{- i \phi/2} 
+ \dots \right] 
\, , \nonumber\\
(3)+(6) 
&= (2) - (5) = (3) - (6) 
= \chi_1^3 \chi_4 \left[  {\rm e}^{- \tau}  {\rm e}^{i \phi/2} - {\rm e}^{-2 \tau} {\rm e}^{\sigma + 3i \phi/2 } + \dots \right]
\, , 
\end{align}
with the rest inducing no effect in the structure in question, $(1)+(4) = (1) - (4) = \chi_1^3 \chi_4 \cdot 0+ \dots$. Here the equality between invariants holds only for 
the $\chi_1^3 \chi_4$ Grassmann component and the ellipses stand for the higher order terms in ${\rm e}^{- \tau}$-expansion.

\subsection{Twist-one contribution}

To start with, let us analyze the twist-one contribution. It arises from the fermion flux-tube excitation. However, to properly describe the NMHV coupling one has to introduce
an additional ad hoc NMHV form factor, given by the power of the Zhukowski variable, such that 
\begin{align}
\label{W1Particle}
W_{\Psi}^{(1)} =  \int_C d \mu_{\Psi} (u) i x [u]
\, ,
\end{align}
where the contour $C$ runs on a two-sheeted Riemann surface glued at the cut $[-2g, 2g]$ on the real axis. It was described in detail in Ref.\ \cite{Basso:2014koa}. According 
to this, the integral splits into two, one going over the large fermion sheet and another over the small one. However, since the semi-circle integration contour for the small 
fermion does not encounter any poles in its interior, the resulting contribution vanishes by Cauchy theorem. This implies that twist-one behavior of the amplitude is governed 
solely by the large fermion
\begin{align}
\label{SingleFermionW}
W_{\Psi}^{(1)} =  \int_{\mathbb{R} + i 0} d \mu_{\rm F} (u) i x [u] 
\, .
\end{align}
Here we used the convention (for ${\rm p} = {\rm F}$)
\begin{align}
d \mu_{\rm p} (u) \equiv \frac{du}{2 \pi} \mu_{\rm p} (u) {\rm e}^{- \tau (E_{\rm p} (u) - 1) + i \sigma p_{\rm p} (u)}
\, ,
\end{align}
where $\mu_{\rm p} (u)$ is a one-particle measure \cite{Basso:2014koa} and $E_{\rm p} (u)$ and $p_{\rm p} (u)$ are its energy and momentum \cite{Basso:2010in}. 
By expanding all functions of the coupling constant in perturbative series, the resulting integrals can be computed using the Cauchy theorem and summing over
the residues  \cite{Basso:2013aha,Papathanasiou:2013uoa,Hatsuda:2014oza,Belitsky:2014sla}. The explicit expression for the lowest two orders reads
\begin{align}
W_{\Psi}^{(1)} 
&= 
\frac{{\rm e}^{- \tau}}{1 + {\rm e}^{2 \sigma}}
\Big\{
g^2
\\
&
-
g^4
\left[
2 \tau
\left(
2 \sigma {\rm e}^{2 \sigma} + (1 - {\rm e}^{2 \sigma}) \ln (1 + {\rm e}^{2 \sigma}) 
\right)
+
(1 - {\rm e}^{2 \sigma}) (2\sigma - \ln (1 + {\rm e}^{2 \sigma})) \ln (1 + {\rm e}^{2 \sigma})
\right]
+
O (g^6)
\Big\}
\, , \nonumber
\end{align}
with further terms being too cumbersome to be quoted here. One can immediately demonstrate that \re{SingleFermionW} agrees\footnote{See the Mathematica notebook 
attached with this submission.} with the recent calculation up to four loops\footnote{The ready-to-use form of the collinear expansion of the NMHV hexagon function $V$ and 
$\widetilde{V}$ is available in Ref.\ \cite{DixonLink}.} \cite{Dixon:2014iba,DixVonHip15}.

At this moment, let us point out that to the twist-two accuracy, that we are currently using, the bosonic Wilson loop $W_6$ can be approximated by its leading 
twist contribution coming from the single gluon flux-tube excitation \cite{Basso:2013aha} 
\begin{align}
W_6 = {\rm e}^{- \tau} \int_{\mathbb{R}} d \mu_{\rm g} (u) + O ({\rm e}^{-2 \tau})
\, . 
\end{align}

\subsection{Twist-two contributions}

Turning to the twist two case, now we encounter two contributions which differ by the total helicity. One of them is accompanied by the ${\rm e}^{3i \phi/2}$ dependence 
and it stems from the fermion--gluon flux-tube states 
\begin{align}
\label{Wfermiongluon}
W_{\rm g {\Psi}}^{(2)} = \int_{\mathbb{R}} d \mu_g (u) \int_C d \mu_{\Psi} (v) \, i x[v] F_{\rm g {\Psi}} (0|u, v) F_{{\Psi} \rm g } (-v, -u | 0)
\, .
\end{align}
The second one possesses the total helicity $- \ft12$ and as a result can come from two distinct two-particle states, hole--antifermion and antigluon--fermion. The latter 
provide additive contributions to the resulting amplitude,
\begin{align}
W_{\rm h {\bar\Psi}/\bar{g} {\Psi}}^{(2)} = W_{\rm h {\bar\Psi}}^{(2)} + W_{\rm \bar{g} {\Psi}}^{(2)} 
\, .
\end{align}
Of course, the hole--antifermion system has to transform in the fundamental representation of SU(4). The individual terms read
\begin{align}
W_{\rm h {\bar\Psi}}^{(2)}
&
= 3 \int_{\mathbb{R}} d \mu_{\rm h} (u) \int_C d \mu_{\Psi} (v) i F^{\bf 4}_{\rm h {\bar\Psi}} (0|u, v) F^{\bf 4}_{{\bar\Psi} \rm h} (- v, - u | 0)
\, , \\
W_{\rm \bar{g} {\Psi}}^{(2)}
&
= \int_{\mathbb{R}} d \mu_{\rm g} (u) \int_C d \mu_{\Psi} (v) \, i x[v] F_{\rm \bar{g} {\Psi}} (0|u, v) F_{{\Psi} \rm \bar{g}} (- v, - u | 0)
\, ,
\end{align}
where the factor of $3$ comes from the SU(4) weight. Notice that in some integrands, we introduced an extra NMHV form factor $x[u]$ for the fermion coupling that is inherited 
from the one-particle contribution \re{W1Particle}.

\subsubsection{Gluon--fermion states}

As we already reviewed above, the integral with respect to the fermion rapidity in Eq.\ \re{Wfermiongluon} goes over a contour $C$ that runs (in a small vicinity) along the 
real axis on the large fermion sheet, then passes through the cut $[-2g, 2g]$ to the small one, where, it encircles an infinite half-circle in the lower semiplane and then 
goes back through the cut on the large fermion sheet to negative infinity \cite{Basso:2014koa}. We observed earlier that the contribution from the small-fermion sheet is 
possible provided the integrand develops a pole inside the integration contour. As we can immediately see from Eq.\ \re{GFandGbarFrelations}, the gluon--small-fermion 
form factor $F_{\rm g {\Psi}} (0|u, v)$ does indeed have a rational prefactor since
\begin{align} 
\frac{1}{P_{\rm g | f} (u|v)} = \frac{\bar{f}_{\rm {\bar g}f} (u, v)}{u - v + \ft{i}{2}} \frac{1}{P_{\rm {\bar g} | f} (u|v)}
\, , 
\end{align}
where the pentagon $P_{\rm {\bar g} | f} (u|v)$ does not possess additional zeroes. Thus, we find for the twist-two term $W_{\rm g{\Psi}}^{(2)}$ a sum of two contributions
\begin{align}
W_{\rm g{\Psi}}^{(2)} = W_{\rm gf}^{(2)} + W_{\rm gF}^{(2)}
\, ,
\end{align}
which read
\begin{align}
W_{\rm gf}^{(2)}
&= \int_{\mathbb{R} + i 0} d \mu_{\rm gf} (u)
\, , \\
\label{gFW}
W_{\rm gF}^{(2)}
&=  \int_{\mathbb{R} + i 0} d \mu_g (u)  \int_{\mathbb{R} + i 0}  d \mu_F (v) \, \frac{i x[v]}{ P_{\rm g | F} (u| v) P_{\rm g | F} (- u| -v)}
\, .
\end{align}
Here in the first equation, the composite measure takes the form
\begin{align}
d \mu_{\rm gf} (u) = \frac{du}{2 \pi}  \mu_{\rm gf} (u) {\rm e}^{- \tau [E_{\rm g} (u) + E_{\rm f} (u^-) - 2] + i \sigma [p_{\rm g} (u) + p_{\rm f} (u^-)]}
\, , 
\end{align}
with
\begin{align}
\mu_{\rm gf} (u)
=
i g^2 \frac{ \mu_{\rm g} (u) \mu_{\rm f} (u^-)}{x[u^-]}\frac{\bar{f}_{\rm {\bar g}f} (u, u^-) \bar{f}_{\rm \bar{g}f} (-u,-u^-)}{P_{\rm {\bar g} | f} (u|u^-) P_{\rm {\bar g} | f} (-u|-u^-)}
\, .
\end{align}
being expressed in terms of the gluon--small-fermion pentagons, one-particle measures \cite{Basso:2013aha,Basso:2014koa}  and an ad hoc form factor continued to 
the small-fermion kinematics.

Counting the powers of the coupling constant in the above two equations, one immediately finds that while $W_{\rm gf}^{(2)}$ starts
at order $g^2$ and thus generates a nonvanishing tree NMHV amplitude, the onset of $W_{\rm gF}^{(2)}$ is postponed to order $g^6$ and therefore
contributes to the NMHV amplitude starting from two loops only. These phenomena were previously observed for MHV and NMHV amplitudes in Refs.\
\cite{Basso:2014koa} and \cite{Belitsky:2014sla}, respectively. They  can immediately be tested making use of the available results for the NMHV 
superamplitude that was recently bootstrapped to four-loop order in Refs.\ \cite{Dixon:2014iba,DixVonHip15}.

The perturbative calculation of integrands is straightforward, to next-to-next-to-leading order the composite gluon--small-fermion measure reads
\begin{align}
\mu_{\rm gf} (u) 
&
= 
\frac{\pi u^-}{i \cosh (\pi u)}
\bigg\{
g^2
\\
&+ g^4
\left[ 
- \ft{1}{2} \left( H_{1/2 - i u} + H_{1/2 + i u} \right)^2 - \frac{\pi  \left( 3 \pi u^- + \sinh (2 \pi u) \right)}{2 u^- \cosh^2 (\pi u)}
-
\frac{i u}{(u^+ u^-)^2} + 5 \zeta_2
\right]
\nonumber\\
&
+ g^6
\bigg[
\frac{1}{2} \left(H_{1/2 - i u} + H_{1/2 + i u} \right) 
\left(
H^{\prime\prime}_{1/2 - i u} + H^{\prime\prime}_{1/2 + i u} 
\right)
+
\frac{1}{4} 
\left(H^{\prime}_{1/2 - i u} + H^{\prime}_{1/2 + i u} \right)^2
\nonumber\\
&\qquad
+
\frac{1}{8} \left(H_{1/2-i u} + H_{1/2 + i u} \right)^4
+ 
\frac{\pi \tanh (\pi u)}{2 u^-} \left(H_{1/2-i u}+H_{1/2 + i u}\right)^2
+
6 \zeta_3 \left( H_{1/2 - i u} + H_{1/2 + i u} \right)
\nonumber\\
&\qquad
-
\frac{ (8 u^2 - 1) (\cosh (2\pi u) + 1) - \pi^2 (u^+ u^-)^2 (5 - \cosh(2\pi u))}{8 (u^+ u^-)^2 \cosh^2 (\pi u)}
\left(H_{1/2-i u} + H_{1/2+i u}\right)^2
\nonumber\\
&\qquad
+
\frac{8 u^2+2 i u-1}{4 (u^+ u^-)^2} \left(H_{1/2-i u} + H_{1/2+i u}\right)^2
-
\frac{\pi ^2  \left(8 u^2 + 3 \pi ^2 (u^+ u^-)^2 - 1 \right)}{2 (u^+ u^-)^2 \cosh^2(\pi  u)}
\nonumber\\
&\qquad
+
\frac{\pi ^2 \tanh (\pi  u)}{16(u^+ u^-)^2}
\left[ 4 \left(8 u^2-1\right) \tanh (\pi  u) + 4 (u^+)^2 {\rm sech}^2 (\pi u) (\sinh (2 \pi  u)+10 \pi u^- )\right]
\nonumber\\
&\qquad
+
\frac{2 \pi \sinh (2 \pi  u) [1 -4 u^2-2 \pi ^2 (u^+ u^-)^2] +\pi^2  [2 u +4 u^2 (26 u-5 i)+7 i]}{16 (u^+)^2 (u^-)^3 \cosh^2(\pi  u)}
\nonumber\\
&\qquad
+
\frac{\pi ^2}{96}\frac{128 u^6+48 u^4-1}{(u^+ u^-)^4}
+
\frac{128 u^4 - 80 u^2 + 3}{8(u^+ u^-)^4}
-
\frac{5 + 4 u \, [12 u^2 (5 u+i) - 32 u -3 i]}{16 (u^+ u^-)^4}
\nonumber\\
&\qquad
-
\frac{\pi ^2 ( 22 u^2+7 i u - 3)}{6(u^+ u^-)^2}
+
\frac{ \pi^4}{16} \frac{\tanh^2(\pi  u)}{\cosh^2(\pi u)}  (\cosh (2 \pi u) - 25) 
+
\frac{9 \pi^4}{4 \cosh^4(\pi  u)}
-
\frac{\pi ^4}{60}
\bigg]
\nonumber\\
&
+ \dots
\bigg\}
\, , \nonumber
\end{align}
where we used harmonic numbers $H_u  = \psi (u+1) + \gamma_{\rm E}$, re-expressed in terms of digamma function $\psi (u) = d \ln \Gamma (u)/du$, and their 
derivatives $H_u^\prime = dH_u/du$ etc. For the gluon--large-fermion contribution, we find that the leading term in Eq.\ \re{gFW} starts at $O (g^6)$ and reads
\begin{align}
W_{\rm gF}^{(2)}= g^6 \int_{\mathbb{R} + i 0} \frac{du}{2 \pi}  \int_{\mathbb{R} + i 0} 
\frac{dv}{2 \pi} {\rm e}^{2i \sigma (u + v)} \frac{i \pi^3}{v (u^2 + \ft{1}{4})} \frac{\tanh(\pi u) - \coth (\pi v)}{ \cosh(\pi u) \sinh (\pi v)}
+
O (g^8)
\, .
\end{align}
The lowest two orders, i.e., tree and one loop, were checked analytically against hexagon bootstrap prediction of Ref.\ \cite{Dixon:2011nj}. They read
\begin{align}
W_{\rm g{\Psi}}^{(2)} 
&
=
\frac{{\rm e}^{- 2 \tau + \sigma}}{(1 + {\rm e}^{2 \sigma})^2}
\Big\{
- g^2
+
g^4
\Big[
2 \tau 
\left(
1 - (1 + 4 \sigma) {\rm e}^{2 \sigma} - 2 \sigma {\rm e}^{4 \sigma}
-
(1 - 2 {\rm e}^{2 \sigma} - {\rm e}^{4 \sigma}) \ln \left(1 + {\rm e}^{2 \sigma} \right)
\right)
\nonumber\\
&
+
2 \sigma (1 - {\rm e}^{2 \sigma}  ) 
-
 2
\left(
1 + \sigma
- 
{\rm e}^{2 \sigma } (1 + 2 \sigma) 
- 
{\rm e}^{4 \sigma} \sigma 
\right)
\ln \left(1 + {\rm e}^{2 \sigma}\right)
+
\left(1 - 2 {\rm e}^{2 \sigma} - {\rm e}^{4 \sigma} \right) \ln^2 \left(1 + {\rm e}^{2 \sigma} \right) 
\Big]
\nonumber\\
&\qquad\qquad\qquad\ 
+
O(g^6)
\Big\}
\, . 
\end{align}
While the two-, three- and four-loop agreement was established numerically to a high precision, confirming the correctness of the pentagon form factors involved 
(see the ancillary file).

\subsubsection{Hole--antifermion and antigluon--fermion states}

Now we are in a position to discuss the last twist-two contribution to the Grassmann component in question. Starting with the hole--antifermion form factor \re{HoleAntiFff4}, 
we immediately conclude that there is a pole in it that induces a nonvanishing effect from the small-fermion sheet of the Riemann surface. Thus again, the contribution gets 
decomposed into two 
\begin{align}
W_{\rm h {\bar\Psi}}^{(2)} = W_{\rm h\bar{f}}^{(2)} + W_{\rm h\bar{F}}^{(2)}
\, , 
\end{align}
where
\begin{align}
W_{{\rm h\bar{f}}}^{(2)} 
&
= \int_{\mathbb{R} + i0} d \mu_{\rm hf} (u)
\, , \\
W_{{\rm h\bar{F}}}^{(2)} 
&
=
3
\int_{\mathbb{R} + i0} d \mu_{\rm h} (u) \int_{\mathbb{R} + i0}  d \mu_{\rm F} (v) \, 
\frac{i}{[(u - v)^2 + \ft94 ] P_{\rm h | F} (u|v) P_{\rm h | F} (-u|-v)}
\, .
\end{align}
Here the composite measure
\begin{align}
d \mu_{\rm hf} (u) 
=  
\frac{du}{2 \pi} \mu_{\rm hf} (u) {\rm e}^{ - \tau \left[ E_{\rm h} (u) + E_{\rm } (u - \frac{3i}{2}) - 2 \right] + i \sigma \left[ p_{\rm h} (u) + p_{\rm f} (u - \frac{3i}{2}) \right]}
\, , 
\end{align}
with
\begin{align}
\mu_{\rm hf} (u) = \frac{i g^2 \mu_{\rm h} (u) \mu_{\rm f} (u - \ft{3 i}{2})}{P_{\rm h | f} \left(u|u - \ft{3i}{2}\right) P_{\rm h | f} \left(-u|-u + \ft{3i}{2} \right)}
\, ,
\end{align}
starts at order $g^2$ in perturbation theory. Finally, the integrand of the antigluon--fermion contribution $W_{\rm \bar{g} {\Psi}}^{(2)}$ does not possess poles on 
the small-fermion sheet and thus receives a nontrivial contribution only from the large-fermion state. Hence, we can write
\begin{align}
W_{\rm \bar{g} {\Psi}}^{(2)}
&
= \int_{\mathbb{R} + i 0}  d \mu_{\rm g} (u) \int_{\mathbb{R} + i 0} d \mu_{\rm F} (v) 
\frac{i x[v]}{P_{\rm {\bar g} | F} (u|v) P_{\rm {\bar g} | F} (-u|-v)}
\, .
\end{align}

The expansion of these twist-two formulas in 't Hooft coupling can be performed to any loop order and then integrals computed numerically. The small antifermion
sets in the earliest in the perturbative expansion inducing, therefore, the tree-level NMHV amplitude. The corresponding effective measure is,
\begin{align}
\mu_{\rm hf} (u) 
&
=
\frac{\pi \left(u - \ft{3i}{2}\right)}{i \cosh(\pi u)}
\bigg\{ g^2
\\
&
+ g^4
\bigg[
- H_{-1/2-i u}^2 - H_{-1/2+i u}^2 + 2 \zeta_2 (1 - 3 \, {\rm sech}^2 (\pi u)) - \frac{\pi \tanh (\pi u)}{u - \ft{3 i}{2}}
\bigg]
\nonumber\\
&
+ g^6
\bigg[
H_{-1/2 -i u} H^{\prime\prime} _{-1/2 - i u} + H_{-1/2 + i u} H^{\prime\prime} _{-1/2 + i u}
+
\frac12 \left( H^\prime_{-1/2 - iu} \right)^2 + \frac12 \left( H^\prime_{-1/2 + iu} \right)^2
\nonumber\\
&\qquad
+
\frac12 \left(H_{-1/2 - i u}^2 + H_{-1/2 + iu}^2 \right)^2
+
6 \zeta_3 \left( H_{-1/2 - i u} + H_{-1/2 + i u} \right)
+
\frac{1}{\left(u - \ft{3 i}{2}\right)^4} 
+ 
\frac{\zeta_2 }{\left(u - \ft{3 i}{2}\right)^2} 
\nonumber\\
&\qquad
+
\frac{\pi  \tanh (\pi  u)}{u-\ft{3 i}{2}}
\left[ H_{-1/2 - i u}^2 + H_{-1/2+i u}^2+ 2 \zeta_2 \left(3 \, {\rm sech}^2(\pi u)-1 \right) \right]
\nonumber\\
&\qquad
+
\frac{6 \zeta_2}{\cosh^2 (\pi u)} \left(H_{-1/2 - i u}^2 + H_{-1/2 + iu}^2 \right)
+
\frac{\pi^4}{2 \cosh^2 (\pi u)}  \left(4 \, {\rm sech}^2 (\pi u) - 3 \right)
\nonumber\\
&\qquad
-
\frac{\pi  \tanh (\pi u)}{\left(u-\ft{3 i}{2}\right)^3}
+
\frac{2 \pi \zeta_2  \tanh (\pi  u)}{\left(u-\ft{3 i}{2}\right)^2}(1 + 3\, {\rm sech}^2(\pi u)) 
-
\frac{\pi^4}{60}
\bigg]
\nonumber\\
&
+
\dots
\bigg\}
\, ,
\nonumber
\end{align}
where the ellipses stand for higher order terms in the perturbative expansion which are too cumbersome to be presented here. The above hole--small-fermion
state generates the tree and one-loop NMHV amplitude,
\begin{align}
W_{{\rm h\bar{f}}}^{(2)} 
&
=
\frac{{\rm e}^{- 2 \tau + \sigma}}{(1 + {\rm e}^{2 \sigma})^2}
\Big\{
- g^2 (2 + {\rm e}^{2 \sigma})
+
g^4
\Big[
2 \tau 
\left(
1 + 4 \sigma - (1 - 2 \sigma) {\rm e}^{2 \sigma} 
-
2 (2 + {\rm e}^{2 \sigma} ) \ln \left(1 + {\rm e}^{2 \sigma} \right)
\right)
\nonumber\\
&
+
2 \sigma (1 - {\rm e}^{2 \sigma}  ) 
-
 2
\left(
1 + 4 \sigma
- 
{\rm e}^{2 \sigma } (1 - 2 \sigma) 
\right)
\ln \left(1 + {\rm e}^{2 \sigma}\right)
+
2
\left(2 + {\rm e}^{2 \sigma} \right) \ln^2 \left(1 + {\rm e}^{2 \sigma} \right) 
\Big]
\nonumber\\
&\qquad\qquad\qquad\ 
+
O(g^6)
\Big\}
\, . 
\end{align}
At two-loop order, the genuine two-particle states come into play and read
\begin{align}
W_{{\rm h\bar{F}}}^{(2)}
&
=
g^6
\int_{\mathbb{R} + i0} \frac{d u}{2 \pi} \int_{\mathbb{R} + i0}  \frac{dv}{2 \pi} \,
{\rm e}^{2i (u+v)\sigma}
\, \frac{3 i \pi^3}{v [(u - v)^2 + \frac{9}{4}] } \frac{\coth (\pi v) - \tanh(\pi u)}{\cosh (\pi u) \sinh (\pi v)}
+ 
O(g^8)
\, , \\
W_{\rm \bar{g} {\Psi}}^{(2)}
&= 
g^6
\int_{\mathbb{R} + i0} \frac{d u}{2 \pi} \int_{\mathbb{R} + i0}  \frac{dv}{2 \pi} \, 
{\rm e}^{2i (u+v)\sigma}
\, \frac{i \pi^3 v}{ (u^2 + \frac{1}{4}) [(u - v)^2 + \frac{1}{4}] } \frac{\coth (\pi v) - \tanh(\pi u) }{\cosh (\pi u) \sinh (\pi v)}
+ 
O(g^8)
\, .
\end{align}
We further computed the next subleading terms in $g$ and numerically verified that the sum of all contributions is indeed in agreement with the three- and
four-loop predictions of Refs.\ \cite{Dixon:2014iba,DixVonHip15}.

\section{Conclusions}

In this paper, within the framework of the pentagon operator product expansion introduced in Ref.\ \cite{Basso:2013vsa}, we discussed twist-two contributions 
to the hexagon NMHV amplitude, elaborating and generalizing our previous consideration \cite{Belitsky:2014sla}. Using a set of fundamental axioms for pentagon 
transitions with nonvanishing R-change, we constructed their nonperturbative solutions and then used known mirror transformations for flux-tube excitations in 
order to find form factors which define their coupling to the Wilson loop contour. These then withstood perturbative tests against the near-collinear limit of 
recent multiloop calculations \cite{Dixon:2014iba,DixVonHip15}.

This study demonstrated, echoing the analysis of the NMHV gluonic component in Ref.\ \cite{Basso:2013aha}, that in addition to pentagon form factors determined
by the bootstrap axioms, NMHV components of the super Wilson loop require introduction of ad hoc form factors which are given in terms of powers of the 
Zhukowski variable, like for the ${\bf 6}$ channel of NMHV amplitude \cite{Belitsky:2014sla}. Currently the form of these extra ingredients do not appear to be driven by 
any fundamental principles and have to be introduced  to achieve agreement with available NMHV data. A proper understanding of this question begs for further 
exploration. We also hope that current results will help to unravel to superstructure of the super Wilson loop and pinpoint a way to its superspace formulation, if it 
exists. Notice, however, that flux-tube excitations break supersymmetry beyond leading order as can be easily verified from the form of their dispersion relations.

Current computing power restricts one's ability to go to even higher orders in perturbation theory within the hexagon bootstrap program 
\cite{Dixon:2013eka,Dixon:2014voa,Dixon:2011nj,Dixon:2014iba}, though there are successful efforts under way to reach four-loop NMHV six-point 
amplitude \cite{DixVonHip15}. As a next step, it is natural to turn to higher point scattering amplitudes, with heptagons coming into the focus. While the OPE data 
can be eagerly provided as a boundary condition for generalizations of the bootstrap approach similar to the one adopted for the hexagon, the lack of a global function 
space is the main obstacle in its immediate implementation. A progress along these line had been achieved at two-loop order in Ref.\ \cite{Golden:2014xqa}.

\section*{Acknowledgments}

We would like to thank Lance Dixon for providing results of Refs.\ \cite{Dixon:2014iba,DixVonHip15} prior to publication and very instructive discussions, and together 
with Matt von Hippel for help in cross-checking numerical data. This work was supported by the U.S. National Science Foundation under the grant PHY-1068286
and PHY-1403891.

\appendix

\setcounter{section}{0}
\setcounter{equation}{0}
\renewcommand{\theequation}{\Alph{section}.\arabic{equation}}

\section{Explicit S-matices}
\label{PTforfs}

In this appendix we give a summary of scattering matrices for flux-tube excitations as well as their mirror transformed images. Since different representations
are advantageous for different purposes, we start this appendix with a universal form of flux-tube phases in terms of sources of the flux-tube
equations \cite{Basso:2013aha}. These expressions are particularly suited for perturbative expansion of S-matrices and, as a consequence, pentagons at 
small value of the 't Hooft coupling. The functions defining scattering amplitudes can be rewritten in the form\footnote{Due to a difference in the definition of 
the sources, our sum representation slightly deviates from analogous formulas in Ref.\ \cite{Basso:2013aha}.}
\begin{align}
f^{(1)}_{{\rm pp}^\prime} (u, v) 
&
= - 2 \widetilde\kappa^{\rm p}_n (u) n \left[ \delta_{nm} - K_{nm} + K_{nl} K_{lm} -\dots \right] \kappa^{{\rm p}^\prime}_m (v)
\, , \\
f^{(2)}_{{\rm pp}^\prime} (u, v) 
&
= - 2 \kappa^{\rm p}_n (u) n \left[ \delta_{nm} - K_{nm} + K_{nl} K_{lm} -\dots \right] (-1)^m \widetilde\kappa^{{\rm p}^\prime}_m (v)
\, , \\
f^{(3)}_{{\rm pp}^\prime} (u, v) 
&
= - 2 \widetilde\kappa^{\rm p}_n (u) n \left[  \delta_{nm} - K_{nm} + K_{nl} K_{lm} -\dots \right] (-1)^m \widetilde\kappa^{{\rm p}^\prime}_m (v)
\, , \\
f^{(4)}_{{\rm pp}^\prime} (u, v) 
&
= - 2 \kappa^{\rm p}_n (u) n (-1)^n \left[  \delta_{nm} - K_{nm} + K_{nl} K_{lm} -\dots \right]  \kappa^{{\rm p}^\prime}_m (v)
\, , 
\end{align}
where the repeated indices are assumed to be summed up from 1 to infinity. Here the coupling-dependent $K$-matrix admits the following integral
representation that can be easily expanded in $g$-series
\begin{align}
K_{nm} 
&
= 2 m (-1)^{m (n +1)} \int_0^\infty \frac{dt}{t} \frac{J_n (2gt) J_{m} (2gt)}{{\rm e}^t - 1}
\\
&
=
2 m (-1)^{m (n +1)} \sum_{k, l = 0}^\infty g^{2k + 2l + n + m} \frac{(-1)^{k + l}(2k + 2l + n + m -1)! \zeta_{2k+2l+n+m}}{k! l! (k+m)! (l+m)!}
\, ,
\end{align}
and the sources $\kappa$ and $\widetilde\kappa$ can be read off from Section 2 of Ref.\ \cite{Belitsky:2014sla}. We do not display them here for the sake of brevity.

Below, we will provide, however, an equivalent representation of the flux-tube phases that explicitly involve the flux-tube solutions $\gamma$, $\widetilde\gamma$. 
This form is particularly useful for studies of the analytic continuation either to the mirror kinematics or from the small- to large-fermion sheet and back. Notice that for a generic 
S-matrix $S_{\rm pp^\prime}$ with ${\rm p} \neq {\rm p}^\prime$, we will quote only one representation in terms of  $\gamma$ and $\widetilde\gamma$ since the other 
ones can easily be recovered by virtue of exchange relations, see \cite{Basso:2013pxa} and in particular Ref.\ \cite{Belitsky:2014sla} for all cases in notations adopted 
in this paper.

\subsection{Hole--(anti)fermion S-matrices}
\label{ExplicitHFfSmatrices}

The S-matrices for hole, and large $\Psi = {\rm F}$ and small $\Psi = {\rm f}$ fermions were discussed previously in Refs.\ \cite{Basso:2014koa,Belitsky:2014sla} 
fermions and read\footnote{And in a different form in Ref.\ \cite{Fioravanti:2013eia}.}, respectively,
\begin{align}
\label{Sphipsi}
S_{\rm hF} (u, v)
&
=
\frac{\Gamma (\ft12 - i u) \Gamma (1 + i v) \Gamma (\ft12 + i u - i v)}{\Gamma (\ft12 + i u) \Gamma (1 - i v) \Gamma (\ft12 - i u + i v)}
\\
&\times\exp\left( 2i \Phi_{\rm hF} (u, v) - 2 i f^{(1)}_{\rm hF} (u, v) + 2 i f^{(2)}_{\rm hF} (u, v) \right)
\, , \nonumber\\
S_{\rm hf} (u, v)
&
=
\exp\left( - 2 i f^{(1)}_{\rm hf} (u, v) + 2 i f^{(2)}_{\rm hf} (u, v) \right)
\, .
\end{align}
Here the exact phase is 
\begin{align}
\Phi_{\rm hF} (u, v)
&=
\int_0^\infty \frac{dt}{t ({\rm e}^t - 1)}
\left(
J_0 (2gt) - 1
\right)
\left(
{\rm e}^{t/2} \sin (ut) - \sin (vt)
\right)
\, , 
\end{align}
and the ones depending on the solutions to the flux-tube equations read for large
\begin{align*}
f^{(1)}_{\rm hF} (u, v) 
&=
\int_0^\infty \frac{dt}{t ({\rm e}^t - 1)} \widetilde\gamma^{\rm h}_{u} (2gt) \left( \cos (vt) - J_0 (2gt) \right)
+ 
\frac{1}{2} \int_0^\infty \frac{dt}{t} \widetilde\gamma^{\rm h}_{+, u} (2gt) \cos (vt)
\, , \\
f^{(2)}_{\rm hF} (u, v) 
&=
\int_0^\infty \frac{dt}{t ({\rm e}^t - 1)} \gamma^{\rm h}_{u} (2gt) \sin (vt)
+
\frac{1}{2} \int_0^\infty \frac{dt}{t} \gamma^{\rm h}_{-, u} (2gt) \sin (vt)
\, , \end{align*}
and small fermions
\begin{align}
f^{(1)}_{\rm hf} (u, v) 
&=
- \frac{1}{2} \int_0^\infty \frac{dt}{t} \widetilde\gamma^{\rm h}_{+, u} (2gt) \cos (vt)
\, , \\
f^{(2)}_{\rm hf} (u, v) 
&=
- \frac{1}{2} \int_0^\infty \frac{dt}{t} \gamma^{\rm h}_{-, u} (2gt) \sin (vt)
\, ,
\end{align}
respectively.

\subsubsection{Mirror in hole rapidity}

Mirror transformation in the hole rapidity immediately yields for above expressions
\begin{align}
\label{SphipsiLargeMirror}
S_{\rm \ast hF} (u, v)
\equiv
S_{\rm hF} (u^\gamma, v)
&
=
\frac{- g^2}{x[v] (u -  v + \ft{i}{2})}
\frac{\Gamma (\ft12 - i u) \Gamma (\ft12 + i u) \Gamma (1 - i v) \Gamma (1 + i v)}{\Gamma (\ft12 - i u + i v) \Gamma (\ft12 + i u - i v)}
\\
&\times\exp\left( 2 \widehat\Phi_{\rm hF} (u, v) + 2 f^{(3)}_{\rm hF} (u, v) - 2 f^{(4)}_{\rm hF} (u, v) \right)
\, , \nonumber\\
\label{SphipsiSmallMirror}
S_{\rm \ast hf} (u, v)
\equiv
S_{\rm hf} (u^\gamma, v)
&
=
\frac{- x[v]}{(u -  v + \ft{i}{2})}
\exp\left( 2 f^{(3)}_{\rm hf} (u, v) - 2 f^{(4)}_{\rm hf} (u, v) \right)
\, .
\end{align}
with the phase
\begin{align}
\widehat\Phi_{\rm hF} (u, v)
&=
\int_0^\infty \frac{dt}{t ({\rm e}^t - 1)}
\left(
J_0 (2gt) - 1
\right)
\left(
{\rm e}^{t/2} \cos (ut) + \cos (vt) - J_0 (2gt) - 1
\right)
\, , 
\end{align}
and the flux-tube functions being
\begin{align*}
f^{(3)}_{\rm hF} (u, v) 
&=
\int_0^\infty \frac{dt}{t ({\rm e}^t - 1)} \widetilde\gamma^{\rm h}_{u} (- 2gt) \sin (vt)
-
\frac{1}{2} \int_0^\infty \frac{dt}{t} \widetilde\gamma^{\rm h}_{-, u} (2gt) \sin (vt)
\, , \\
f^{(4)}_{\rm hF} (u, v) 
&=
\int_0^\infty \frac{dt}{t ({\rm e}^t - 1)} \gamma^{\rm h}_{u} (- 2gt) \left( \cos (vt) - J_0 (2gt) \right)
+ 
\frac{1}{2} \int_0^\infty \frac{dt}{t} \gamma^{\rm h}_{+, u} (2gt) \cos (vt)
\, ,
\end{align*}
for large and 
\begin{align}
f^{(3)}_{\rm hf} (u, v) 
&=
+ \frac{1}{2} \int_0^\infty \frac{dt}{t} \widetilde\gamma^{\rm h}_{-, u} (2gt) \sin (vt)
\, , \\
f^{(4)}_{\rm hf} (u, v) 
&=
- \frac{1}{2} \int_0^\infty \frac{dt}{t} \gamma^{\rm h}_{+, u} (2gt) \cos (vt)
\, ,
\end{align}
for small fermions, respectively.

\subsubsection{Mirror in fermion rapidity}
\label{HoleFermionMirrorS}

Using explicit formulas \re{SmirrorPsih} and the mirror hole--hole S-matrix \cite{Basso:2013pxa}
\begin{align}
S_{\ast\rm{hh}} (u, v)
&
=
g^2
\frac{u - v}{u - v + i}
\frac{\Gamma (\ft12 - i u) \Gamma (\ft12 + i u) \Gamma (\ft12 - i v) \Gamma (\ft12 + i v)}{\Gamma (\ft12 - i u + i v) \Gamma (\ft12 + i u - i v)}
\\
&\times\exp \left( 2 \widehat\Phi_{\rm hh} (u, v) + 2 f^{(3)}_{\rm hh} (u, v) - 2 f^{(4)}_{\rm hh} (u, v) \right)
\, , \nonumber
\end{align}
where all of its ingredients are defined by
\begin{align}
\label{SigmahhAst}
\widehat\Phi_{\rm hh} (u, v)
&=
\int_0^\infty \frac{dt}{t ({\rm e}^t - 1)} \left( J_0 (2gt) - 1 \right)
\left(
{\rm e}^{t/2} \left( \cos(ut) + \cos(vt) \right) - J_0 (2gt) - 1
\right)
\, , \\
f^{(3)}_{\rm hh} (u, v) 
&= 
\int_0^\infty \frac{dt}{t ({\rm e}^t - 1)} {\rm e}^{t/2} \sin(ut) \widetilde\gamma^{\rm h}_v (- 2gt)
\, , \\
f^{(4)}_{\rm hh} (u, v)
&=
\int_0^\infty \frac{dt}{t ({\rm e}^t - 1)}
\left( {\rm e}^{t/2} \cos(ut) - J_0 (2gt) \right) \gamma^{\rm h}_v (- 2gt)
\, ,
\end{align}
we can immediately find the mirror S-matrices for fermions living on the small and large sheets,
\begin{align}
\label{SMirrorfh}
S_{\ast\rm{\bar{f} h}} (u, v) 
&
= \frac{x[u]}{u - v + \ft{i}{2}} \exp \left( 2 f^{(3)}_{\rm fh} (u, v) - 2 f^{(4)}_{\rm fh} (u, v) \right)
\, , \\
S_{\ast\rm{\bar{F}h}} (u, v) 
&
=
\frac{g^2}{x[v] (u -  v + \ft{i}{2})}
\frac{\Gamma (\ft12 - i u) \Gamma (\ft12 + i u) \Gamma (1 - i v) \Gamma (1 + i v)}{\Gamma (\ft12 - i u + i v) \Gamma (\ft12 + i u - i v)}
\\
&\times\exp\left( 2 \widehat\Phi_{\rm Fh} (u, v) + 2 f^{(3)}_{\rm Fh} (u, v) - 2 f^{(4)}_{\rm Fh} (u, v) \right)
\, , \nonumber
\end{align}
with the flux-tube phases reading for small
\begin{align}
f^{(3)}_{\rm fh} (u, v)
&= + \frac{1}{2} \int_0^\infty \frac{dt}{t} \widetilde\gamma^{\rm h}_{-, v} (2gt) \sin (ut)
\, , \\
f^{(4)}_{\rm fh} (u, v)
&= - \frac{1}{2} \int_0^\infty \frac{dt}{t} \gamma^{\rm h}_{+, v} (2gt) \cos (ut)
\, , 
\end{align}
and large fermions
\begin{align}
f^{(3)}_{\rm Fh} (u, v) 
&=
\int_0^\infty \frac{dt}{t ({\rm e}^t - 1)} \widetilde\gamma^{\rm h}_{v} (- 2gt) \sin (ut)
-
\frac{1}{2} \int_0^\infty \frac{dt}{t} \widetilde\gamma^{\rm h}_{-, v} (2gt) \sin (ut)
\, , \\
f^{(4)}_{\rm Fh} (u, v) 
&=
\int_0^\infty \frac{dt}{t ({\rm e}^t - 1)} \gamma^{\rm h}_{v} (- 2gt) \left( \cos (ut) - J_0 (2gt) \right)
+ 
\frac{1}{2} \int_0^\infty \frac{dt}{t} \gamma^{\rm h}_{+, v} (2gt) \cos (ut)
\, ,
\end{align}
respectively. Here, the exact phase is
\begin{align}
\widehat\Phi_{\rm Fh} (u, v)
&=
\int_0^\infty \frac{dt}{t ({\rm e}^t - 1)}
\left(
J_0 (2gt) - 1
\right)
\left(
 \cos (ut)  + {\rm e}^{t/2} \cos (vt) - J_0 (2gt) - 1
\right)
\, .
\end{align}
Let us emphasize one more time that in light of the relations \re{HolePsitoHolePsiBar} and \re{SpsiphiRelation}, the mirror S-matrices found in this 
appendix are related to the previous ones with the mirror hole rapidity.

\subsection{(Anti)gluon--(anti)fermion S-matrices}
\label{GaugeFermionSmatrix}

The large/small-fermion--gluon S-matrices are (${\Psi} = {\rm F, f}$)  \cite{Belitsky:2014sla} 
\begin{align}
S_{\rm gF} (u, v) 
&
= 
\frac{\Gamma (\ft32 - i u) \Gamma (1 + i v) \Gamma (\ft32 + i u - i v)}{\Gamma (\ft32 + i u) \Gamma (1 - i v) \Gamma (\ft32 - i u + i v)}
\\
&\times\exp
\left(
2i \Phi_{\rm gF} (u, v) - 2i f^{(1)}_{\rm gF} (u, v) + 2i f^{(2)}_{\rm gF} (u, v)
\right)
\, , \nonumber\\
\label{Sgf}
S_{\rm gf} (u, v) 
&
= 
\exp\left( 
- 2i f^{(1)}_{\rm gf} (u, v) + 2i f^{(2)}_{\rm gf} (u, v)
\right)
\, ,
\end{align}
where the exact phase is
\begin{align}
 \Phi_{\rm gF}  (u, v)
=
\int_0^\infty \frac{dt}{t ({\rm e}^t - 1)}
\left( J_0 (2gt) - 1 \right)
\left(
{\rm e}^{- t/2} \sin (ut) - \sin (vt) 
\right)
\, ,
\end{align}
and the flux-tube scattering phases are
\begin{align}
f^{(1)}_{\rm gF} (u, v) 
&
=
\int_0^\infty \frac{dt}{t} \frac{\cos (vt) - J_0 (2gt)}{{\rm e}^t - 1} \widetilde\gamma_u^{\rm g} (2gt)
+
\frac{1}{2}
\int_0^\infty \frac{dt}{t} \cos (vt) \widetilde\gamma_{+,u}^{\rm g} (2gt)
\, , \\
f^{(2)}_{\rm gF} (u, v) 
&
=
\int_0^\infty \frac{dt}{t} \frac{\sin (vt)}{{\rm e}^t - 1} \gamma_u^{\rm g} (2gt)
+
\frac{1}{2}
\int_0^\infty \frac{dt}{t} \sin (vt) \gamma_{-,u}^{\rm g} (2gt)
\, .
\end{align}
Moving to the small fermion sheet, we find
\begin{align}
f^{(1)}_{\rm gf} (u, v) 
&
=
-
\frac{1}{2}
\int_0^\infty \frac{dt}{t} \cos (vt) \widetilde\gamma_{+, u}^{\rm g} (2gt)
\, , \\
f^{(2)}_{\rm gf} (u, v) 
&
=
- \frac{1}{2}
\int_0^\infty \frac{dt}{t} \sin (vt) \gamma_{-, u}^{\rm g} (2gt)
\, .
\end{align}
An equivalent representation can be found in Ref.\ \cite{Fioravanti:2013eia}. 

\subsubsection{Mirror in gluon rapidity}

The mirror transformation in gluon rapidity immediately yields
\begin{align}
S_{\ast {\rm gF}} (u, v) 
\equiv
S_{\rm gF} (u^\gamma, v) 
&
= 
- \frac{g^2}{x[v] (u - v + \ft{i}{2})}
\frac{\Gamma (\ft12 - i u) \Gamma (\ft12 + i u) \Gamma (1 - i v) \Gamma (1 + i v)}{\Gamma (\ft12 - i u + i v) \Gamma (\ft12 + i u - i v)}
\\
&
\times
\exp\left( 
2 \widehat\Phi_{\rm gF} (u, v) + 2 f^{(3)}_{\rm gF} (u, v) - 2 f^{(4)}_{\rm gF} (u, v)
\right)
\, , \nonumber\\
\label{Sgfmirror}
S_{\rm \ast gf} (u, v) 
\equiv
S_{\rm gf} (u^\gamma, v) 
&
= 
- \frac{x[v]}{u - v + \ft{i}{2}}
\exp\left( 
2 f^{(3)}_{\rm gf} (u, v) - 2 f^{(4)}_{\rm gf} (u, v)
\right)
\, ,
\end{align}
with
\begin{align}
\widehat\Phi_{\rm gF}  (u, v) 
=
\int_0^\infty
\frac{dt}{t ({\rm e}^t - 1)}
\left(J_0 (2gt) - 1\right)
\left(
{\rm e}^{t/2} \cos (ut) + \cos (vt)  - J_0 (2gt) - 1
\right)
\, ,
\end{align}
and
\begin{align}
f^{(3)}_{\rm gF} (u, v) 
&
=
\int_0^\infty \frac{dt}{t} \frac{\widetilde\gamma^{\rm g}_u (- 2gt)}{{\rm e}^t - 1} \sin(vt) 
- 
\frac{1}{2}
\int_0^\infty \frac{dt}{t} \widetilde\gamma^{\rm g}_{-,u} (2gt) \sin(vt) 
\, , \\
f^{(4)}_{\rm gF}  (u, v) 
&
=
\int_0^\infty \frac{dt}{t} \frac{\gamma^{\rm g}_u (- 2gt)}{{\rm e}^t - 1} \left( \cos(vt) - J_0 (2gt) \right)
+
\frac{1}{2}
\int_0^\infty \frac{dt}{t} \gamma^{\rm g}_{+,u} (2gt) \cos(vt) 
\, , 
\end{align}
and for large fermion and
\begin{align}
f^{(3)}_{\rm gf} (u, v) 
&
= +
\frac{1}{2}
\int_0^\infty \frac{dt}{t} \widetilde\gamma^{\rm g}_{-,u} (2gt) \sin(vt) 
\, , \\
f^{(4)}_{\rm gf} (u, v) 
&
=
-
\frac{1}{2}
\int_0^\infty \frac{dt}{t} \gamma^{\rm g}_{+,u} (2gt) \cos(vt) 
\, ,
\end{align}
for the small one.

\subsubsection{Mirror in fermion rapidity}
\label{GaugeFermionMirror}

Using the defining relation \re{fbargmirrorSmatrix}, we find for small antifermion in the crossed channel
\begin{align}
S_{\ast\rm \bar{f}g} (u, v) = \frac{x[u]}{u - v + \ft{i}{2}} 
\exp \left( 
2 f_{\rm fg}^{(3)} (u, v)
-
2 f_{\rm fg}^{(4)} (u, v)
\right)
\, ,
\end{align}
where
\begin{align}
f_{\rm fg}^{(3)} (u, v)
&
=
+ \frac{1}{2} \int_0^\infty \frac{dt}{t} \sin (ut) \widetilde\gamma^{\rm g}_{-, v} (2gt)
\, , \\
f_{\rm fg}^{(4)} (u, v)
&
=
- \frac{1}{2} \int_0^\infty \frac{dt}{t} \cos (ut) \gamma^{\rm g}_{+, v} (2gt)
\, .
\end{align}
Moving the rapidity $u$ to the large fermion sheet, we get
\begin{align}
S_{\ast\rm \bar{F}g} (u, v) 
&
= \frac{g^2}{x[u] (u - v + \ft{i}{2})} 
\frac{ \Gamma (1 - i u) \Gamma (1 + i u) \Gamma (\ft12 - i v) \Gamma (\ft12 + i v)}{\Gamma (\ft12 - i u + i v) \Gamma (\ft12 + i u - i v)}
\\
&\times\exp\left( 2 \widehat\Phi_{\rm Fg} (u, v) + 2 f^{(3)}_{\rm Fg} (u, v) - 2 f^{(4)}_{\rm Fg} (u, v) \right)
\, , \nonumber
\end{align}
where
\begin{align}
\widehat\Phi_{\rm Fg}  (u, v) 
=
\int_0^\infty
\frac{dt}{t ({\rm e}^t - 1)}
\left(J_0 (2gt) - 1\right)
\left(
\cos (ut) + {\rm e}^{t/2} \cos (vt) - J_0 (2gt) - 1
\right)
\, ,
\end{align}
and
\begin{align}
f^{(3)}_{\rm Fg} (u, v) 
&
=
\int_0^\infty \frac{dt}{t} \frac{\widetilde\gamma^{\rm g}_v (- 2gt)}{{\rm e}^t - 1} \sin(ut) 
- 
\frac{1}{2}
\int_0^\infty \frac{dt}{t} \widetilde\gamma^{\rm g}_{-,v} (2gt) \sin(ut) 
\, , \\
f^{(4)}_{\rm Fg}  (u, v) 
&
=
\int_0^\infty \frac{dt}{t} \frac{\gamma^{\rm g}_v (- 2gt)}{{\rm e}^t - 1} \left( \cos(ut) - J_0 (2gt) \right)
+
\frac{1}{2}
\int_0^\infty \frac{dt}{t} \gamma^{\rm g}_{+,v} (2gt) \cos(ut) 
\, .
\end{align}

 %%%%%%  Bibliography %%%%%%%%%%%%%%%%%%%%%%%%%%%%%%%%%%%%%%%%%%%%

%%%%%%%%%%%%%%%%%%%%%%%%%%%%%%%%%%%%%%%%%%%%%%%%%%%%%%%%%%%%%%%%
\end{document}